\pdfoutput=1
\documentclass[]{aastex631}
\usepackage{graphicx}
\usepackage{amsmath}
\usepackage{epstopdf}

\begin{document}
\title{proto-Lightspeed: a high-speed, ultra-low read noise imager on the Magellan Clay Telescope}
\altaffiliation{This paper includes data gathered with the 6.5 meter Magellan Telescopes located at Las Campanas Observatory, Chile.}

\author[0000-0002-7191-4403]{Christopher Layden}
\altaffiliation{Corresponding author;  \href{mailto:clayden7@mit.edu}{clayden7@mit.edu}}
\affiliation{MIT Kavli Institute for Astrophysics and Space Research, Massachusetts Institute of Technology, 77 Massachusetts Ave, Cambridge, MA 02139, USA}
\affiliation{MIT Department of Physics, 77 Massachusetts Ave., Cambridge, MA 02139, USA}

\author[0000-0002-7226-836X]{Kevin Burdge}
\affiliation{MIT Kavli Institute for Astrophysics and Space Research, Massachusetts Institute of Technology, 77 Massachusetts Ave, Cambridge, MA 02139, USA}
\affiliation{MIT Department of Physics, 77 Massachusetts Ave., Cambridge, MA 02139, USA}

\author[0000-0001-8467-9767]{Gabor Furesz}
\affiliation{MIT Kavli Institute for Astrophysics and Space Research, Massachusetts Institute of Technology, 77 Massachusetts Ave, Cambridge, MA 02139, USA}

\author[0000-0003-1361-985X]{Juliana García-Mejía}
\altaffiliation{51 Pegasi B Fellow, MIT Pappalardo Physics Fellow}
\affiliation{MIT Kavli Institute for Astrophysics and Space Research, Massachusetts Institute of Technology, 77 Massachusetts Ave, Cambridge, MA 02139, USA}
\affiliation{MIT Department of Physics, 77 Massachusetts Ave., Cambridge, MA 02139, USA}
\affiliation{Center for Astrophysics \textbar\ Harvard \& Smithsonian, 60 Garden Street, Cambridge, MA 02138, USA}

\author[0000-0002-6401-778X]{Jack Dinsmore}
\affiliation{Department of Physics, Stanford University, Stanford, CA 94305, USA}
\affiliation{Kavli Institute for Particle Astrophysics and Cosmology, Stanford University, Stanford, CA 94305, USA}

\author[0000-0001-6331-112X]{Geoffrey Mo}
\affiliation{Department of Astronomy, California Institute of Technology, 1216 E California Blvd, Pasadena, CA 91125, USA}
\affiliation{The Observatories of the Carnegie Institution for Science, 813 Santa Barbara St, Pasadena, CA 91101, USA}

\author[0000-0003-0412-9664]{David Osip}
\affiliation{Las Campanas Observatory, Carnegie Institution for Science, Colina el Pino, Casilla 601 La Serena, Chile}

\author[0000-0002-9602-2217]{John J. Piotrowski}
\affiliation{Observatories of the Carnegie Institution of Washington, 813 Santa Barbara St, Pasadena, CA 91101, USA}

\author[0000-0001-6711-3286]{Roger W. Romani}
\affiliation{Department of Physics, Stanford University, Stanford, CA 94305, USA}
\affiliation{Kavli Institute for Particle Astrophysics and Cosmology, Stanford University, Stanford, CA 94305, USA}

\author[0009-0006-7343-3300]{August Berne}
\affiliation{MIT Department of Aeronautics and Astronautics, 125 Massachusetts Ave., Cambridge, MA 02139, USA}

\author[0000-0001-8804-8946]{Deepto Chakrabarty}
\affiliation{MIT Kavli Institute for Astrophysics and Space Research, Massachusetts Institute of Technology, 77 Massachusetts Ave, Cambridge, MA 02139, USA}
\affiliation{MIT Department of Physics, 77 Massachusetts Ave., Cambridge, MA 02139, USA}

\author[0000-0003-4780-4105]{Emma Chickles}
\affiliation{MIT Kavli Institute for Astrophysics and Space Research, Massachusetts Institute of Technology, 77 Massachusetts Ave, Cambridge, MA 02139, USA}
\affiliation{MIT Department of Physics, 77 Massachusetts Ave., Cambridge, MA 02139, USA}

\begin{abstract}
    \noindent proto-Lightspeed is a new instrument that has been commissioned on the Nasmyth East port of the Magellan Clay Telescope at Las Campanas Observatory to deliver high-speed optical imaging with deep sub-electron read noise. Making use of commercial re-imaging lenses and the ORCA-Quest 2 camera from Hamamatsu, proto-Lightspeed images a field $1'$ in diameter at up to $200$~Hz or windowed fields at higher rates, up to $6600$~Hz for a $1.6''\times 1'$ field of view. proto-Lightspeed delivers seeing-limited image quality in the $g'$, $r'$, and $i'$ bands and adjustable magnification for pixel scales between $0.017''-0.050''$. proto-Lightspeed is well suited to studying compact binary systems, exoplanet transits, rapid flaring associated with accretion, periodic optical emission from pulsars, occultations of background stars by small trans-Neptunian Objects, and any other rapidly variable source. proto-Lightspeed will be a P.I. instrument beginning in 2026B, available for use by members of the Magellan Consortium. In this paper, we discuss the design and performance of the instrument, results from its two commissioning runs, and plans for a facility instrument, Lightspeed, to support simultaneous multicolor imaging across a $7'\times4'$ field. 
\end{abstract}

\keywords{instrumentation: photometers; instrumentation: detectors; pulsars; trans-Neptunian Objects; transit photometry; stellar flares}

\section{Introduction}
\label{sec:intro}
Astrophysical phenomena with optical emission that varies on short time scales ($\lesssim 1$~s) are largely inaccessible to traditional astronomical instruments designed for long stares. Existing high-speed imagers built to probe this parameter space have yielded a wealth of knowledge across a wide swath of astrophysics, from the solar system to extragalactic distances. By allowing high-cadence monitoring of background stars to detect occultations by trans-Neptunian objects (TNOs), they have enabled characterization of rings around TNOs \citep{morgado:2023} and provided some of the tightest constraints on the population of km-scale TNOs \citep{zhang:2023}. By providing multicolor transit light curves, they constrained the shapes, sizes, compositions, and evolution of disintegrating bodies in exoplanet systems, offering a view into planetary end stages \citep{bochinski:2015,gansicke:2016}. They have revealed color evolution in M dwarf flares on $\lesssim 1$~s timescales \citep{Kowalski:2016}, providing direct tests of flare emission mechanisms and informing assessments of habitability around such stars \citep{segura:2010,ramsay:2021}.

High-speed imagers have allowed astronomers to identify new populations of compact binary systems, including the first detected a white dwarf pulsar, and to probe accretion phenomena at higher precision than ever before \citep{marsh:2016,kupfer:2020,sanchez:2023}. In combination with simultaneous X-ray observations, they have uncovered essential clues about the geometry and emission mechanisms of outbursting X-ray binaries (XRBs; \citealt{Gandhi2008,Gandhi2010}). They have discovered periodic optical emission at the spin frequencies of well-localized energetic pulsars \citep{barbieri:2019,Ambrosino:2017}. Phase-resolved optical photometry and polarimetry of young pulsars can address longstanding questions about their magnetospheric emission mechanisms \citep{slowikowska:2009}. Additionally, they have provided the tightest upper limits on the brightness of optical counterparts to fast radio bursts \citep[FRBs;][]{kilpatrick:2024}.

Notable among these instruments are ULTRACAM \citep{dhillon:2007} on the New Technology Telescope (NTT), its successor HiPERCAM \citep{dhillon:2021} on the Gran Telescopio Canarias (GTC), and the Caltech HIgh-speed Multi-color camERA \citep[CHIMERA;][]{Harding_2016} on the Hale telescope. Each of these can provide simultaneous multicolor optical imaging at hundreds of frames per second (if windowed to a small region). These instruments are workhorses for providing high-cadence follow-up to newly discovered variable phenomena identified by new large-scale time domain surveys [e.g., the Zwicky Transient Facility \citep{bellm:2019}, the Vera C. Rubin Observatory \citep{ivezic:2019}, and the Transiting Exoplanet Survey Satellite \citep{ricker:2015}]. The twin instruments `Alopeke and Zorro on the Gemini North and South telescopes \citep{10.3389/fspas.2021.716560} were designed for speckle interferometry but can be operated in wide-field mode to also provide dual-channel high-speed photometry. Geiger-mode instruments like Iqueye on the NTT provide little to no spatial resolution but detect and time-tag individual photons with sub-nanosecond precision \citep{naletto:2009}. Table~\ref{tab:instrument_table} gives more information about these modern high-speed imagers.

The existing spatially resolving high-speed imagers employ charge-coupled devices (CCDs), which have for decades been the dominant optical imaging technology. ULTRACAM and HiPERCAM use frame-transfer CCDs, while CHIMERA, `Alopeke, and Zorro use identical frame-transfer electron multiplying CCDs (EMCCDs) that achieve higher full frame readout rates and sub-electron read noise. However, even when using pixel binning and the smallest window sizes, these instruments do not push far beyond frame rates of 1000~Hz. Moreover, operating conventional CCDs at high speeds results in increased read noise. While EMCCDs can deliver sub-electron read noise, they suffer from a multiplicative excess noise factor and clock-induced charge \citep{Harding_2016,robbins:2003,tulloch:2011}. These noise sources in either CCD variety restrict the sensitivity of these instruments when performing high-speed imaging.

\begin{table}[b!]
    \centering
    \begin{tabular}{c|c|c|c|c|c|c|c}
         Instrument  & \multicolumn{2}{c|}{Telescope} & Colors$^*$ & Max Field & \multicolumn{2}{c|}{Max Frame Rate (Hz)} & Min Readout \\
         Name & Name & Diam (m) & & of View & Windowed & Full Field & Noise ($\mathrm{e}^-$)\\
         \hline
         ULTRACAM & NTT & 3.6 & $u',g',r'/i'/z'$ & $6' \times 6'$ & 500 & 0.3 & 3 \\
         HiPERCAM & GTC & 10.4 & $u',g',r',i',z'$ & $2.8' \times 1.4'$ & $>1000$& 2 & 4.5 \\
         CHIMERA & Hale & 5.1 & $u'/g'$, $r'/i'/z'$ & $5'\times5'$ & $>1000$& 26 & $<1$\\
         `Alopeke/Zorro & Gemini N/S & 8.1 & $u'/g'/r'$, $i'/z'$ & $1'$ (diam) & $>1000$ & 26 & $<1$\\
         Iqueye$^\dag$ & NTT & 3.6 & $B/V/R/I$ & $6.1''$ (diam) & N/A & $8 \times 10^6$ & 0 \\
         proto-Lightspeed & Clay & 6.5 & $g'/r'/i'$ & $1'$ (diam) & $>6600$& 200 & 0.3$^\ddag$\\
         Lightspeed (planned) & Clay & 6.5 & $u',g',r',i',z'$ & $7'\times4'$ & $>8000$ & 120 & 0.3$^\ddag$\\
    \end{tabular}
    \caption{Characteristics of existing high-speed astronomical instruments, compared to proto-Lightspeed and the planned Lightspeed instrument. $^*$Commas indicate filters that can be imaged through simultaneously. Slashes indicate filters that can be switched between asynchronously. Instruments may have filter wheels with other options, e.g., for narrow-band imaging. $^\dag$Iqueye does not provide spatial resolution, and the maximum frame rate is given as the maximum driving frequency of its single-photon-counting SPAD detectors. $^\ddag$See Sec.~\ref{sec:read_noise} for discussion of proto-Lightspeed's measured read noise.}
    \label{tab:instrument_table}
\end{table}

Recent advances in Complementary Metal-Oxide-Silicon (CMOS) image sensor technology have opened an opportunity to develop high-speed imagers with significantly higher readout rates and improved noise characteristics relative to CCDs. CCDs typically have one or a few output amplifiers and analog-to-digital converters (ADCs). In most scientific CMOS images sensors, every row or column has its own ADC, allowing for parallelized readout and faster full frame rates. Significant investment into CMOS image sensor technology for commercial electronics has enabled modern CMOS image sensors to deliver quantum efficiency (QE) peaking above $90\%$ and dark current below 0.1~e$^-$/pix/s with very modest cooling requirements; \citep[$T\gtrsim -25^\circ$C;][]{Alarcon_2023,Khandelwal_2024,layden:2025,Karpov_2020,apergis:2025}.

Some CMOS image sensors have also demonstrated deep sub-electron read noise (DSERN; i.e., $<0.5\,\mathrm{e}^-$) by employing very low capacitance sense nodes  \citep{fossum:2016,ma:2022,gallagher:2024}. These DSERN devices offer an order-of-magnitude improvement in read noise compared to typical CCDs, without the drawbacks of EMCCDs. DSERN CMOS image sensors are also approaching the photon counting regime: when the electron-referred read noise drops below $\approx 0.2\,\mathrm{e}^-$, one may round with confidence to the nearest whole number of photoelectrons, effectively eliminating read noise altogether. The lowest-noise commercially available sensor at present, the HWK4123 (housed in the Hamamatsu ORCA-Quest 2 camera), reaches $\approx0.3\,\mathrm{e}^-$ \citep{gallagher:2024,ma:2022} and can begin to take advantage of resolving individual electrons.

Other detector architectures enabling photon counting have been developed, but with significant downsides. Skipper CCDs \citep{tiffenberg:2017} require very slow readout. Geiger-mode avalanche photodiode (APD) devices, including single-photon avalanche photodiodes (SPADs; used in Iqueye) and silicon photomultipliers (SiPMs), offer little to no spatial resolution and small fields, making comparison photometry, discovery science, and multi-object studies challenging. DSERN CMOS image sensors combine advantages of CCDs (large FOV, spatial resolution) and those of Geiger-mode devices (photon counting, time resolution) while avoiding the pitfalls of each (cryogenic cooling, multiplication noise, dead time). For these reasons, DSERN CMOS image sensors, in particular the ORCA-Quest 2, have recently been deployed on ground-based telescopes, including the Subaru Telescope \citep{lucas:2024} and Calar Alto Observatory \citep{roth:2025}, and have been selected for use in the Lazuli Space Observatory \citep{Roy2026}.

Such DSERN CMOS image sensors are particularly attractive for the next generation of high-speed imagers. With their fast full frame rates, they could yield significantly more detections of km-scale TNOs. They could deliver higher precision, higher cadence photometry for exoplanetary and ultracompact binary systems. With kilohertz imaging and photon counting, they could allow for resolution of optical outbursts in XRBs and new detections of periodic optical emission from pulsars. They could also yield significantly more sensitive searches for optical counterparts to FRBs. However, as CMOS image sensors are still a relatively new technology for astronomy, their behavior should be well understood before they are used to provide astronomical data. For example, nonlinear response has been identified in some CMOS image sensors at low signal levels or at high signal levels. Such nonlinearities are believed to be caused by incomplete charge transfer and the variable capacitance of the sense node, respectively (\cite{layden:2025,janesick_i}).

proto-Lightspeed is a new high-speed imager on the 6.5~m Magellan Clay telescope employing the Hamamatsu ORCA-Quest 2 camera. This instrument also provides other capabilities that benefit greatly from DSERN CMOS image sensor technology, including speckle interferometry and narrow-band imaging. It uses commercial off-the-shelf (COTS) components and was designed for rapid development, allowing for timely demonstration of the advantages of this CMOS device and characterization of its behavior. In this paper, we describe proto-Lightspeed's optical and mechanical design (Sec.~\ref{sec:optical_design}, Sec.~\ref{sec:mech_design}), our characterization and calibration of the ORCA-Quest 2 camera (Sec.~\ref{sec:detector}), and the software developed for proto-Lightspeed (Sec.~\ref{sec:software}). We then report on the commissioning of the instrument at the Nasmyth East (NasE) visitor port of the Magellan Clay Telescope at Las Campanas Observatory (LCO) and summarize its first on-sky results (Sec.~\ref{sec:performance}). We share a few early science highlights from the commissioning runs. Finally, we outline plans to develop the full Lightspeed instrument, which will be a five-channel imager with custom optical components (Sec.~\ref{sec:future}).

\section{Optical Design}
\label{sec:optical_design}
\begin{figure}
    \centering
    \includegraphics[width=0.75\linewidth]{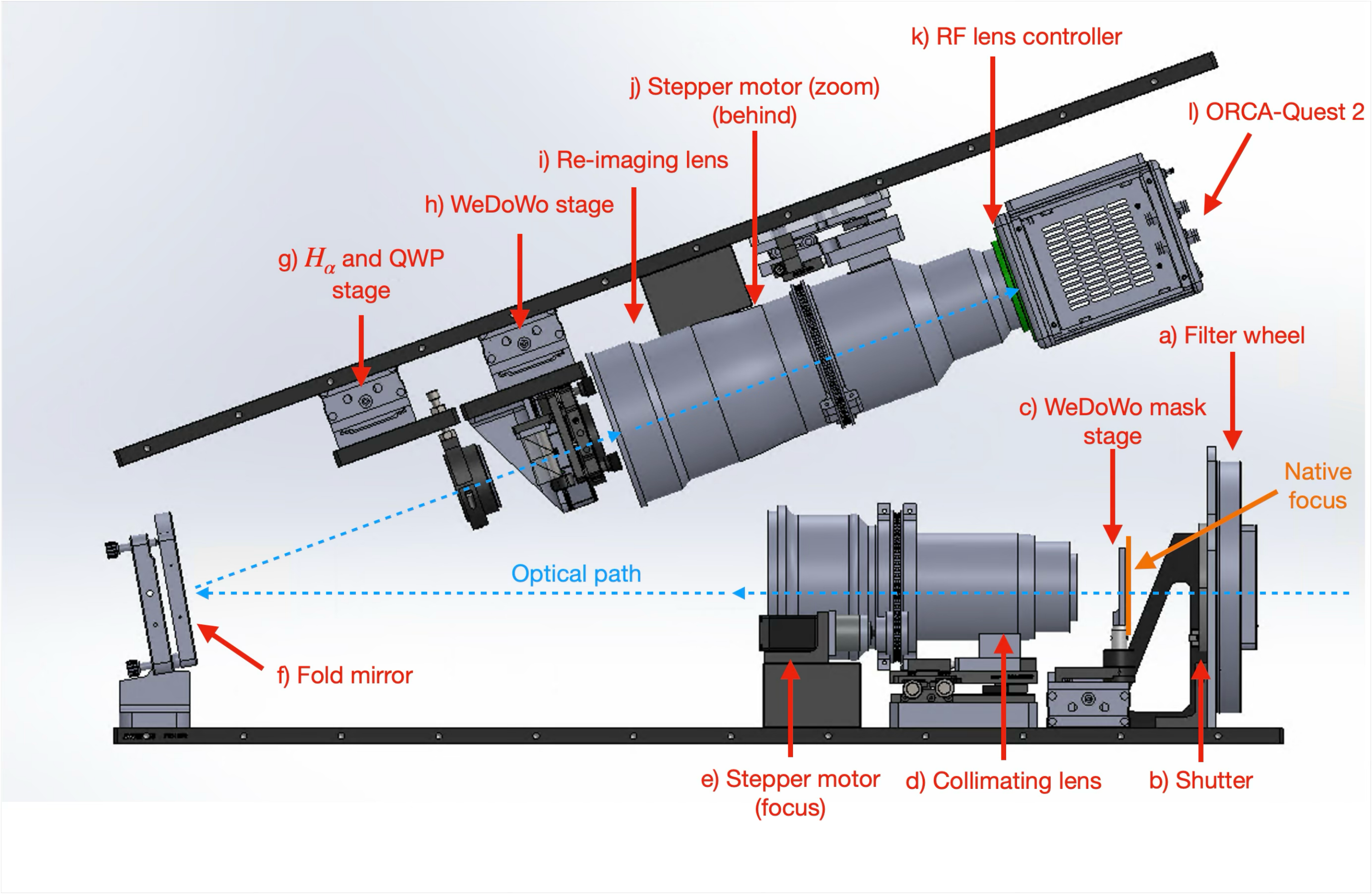}
    \caption{Computer-aided design model of the optical components of proto-Lightspeed. Reflected light from the telescope's tertiary mirror enters the instrument from the right. The length of each breadboard is 36~in.}
    \label{fig:cad_model}
\end{figure}

To achieve a moderate field of view (FOV) and adequate sampling with the ORCA-Quest 2's 4.6~µm pixels, it was necessary to re-image the native focal plane at the NasE port, which is delivered at a focal ratio of $f/11$. The optical design to conduct this re-imaging was selected to meet the following requirements: i) de-magnify the field by a factor of at least 4 to provide a pixel scale of at least $0.05''$/pix (from the native $0.013''$/pix); ii) provide a vignetted FOV at least $1'$ in diameter; iii) provide seeing-limited imaging in the best observing conditions at LCO ($\approx 0.3''$); iv) consist of COTS components to enable inexpensive and rapid deployment. Our design met these requirements and provided the additional capability of variable de-magnification, allowing for switching between speckle imaging and conventional imaging modes. Figure~\ref{fig:cad_model} shows a computer-aided design (CAD) model of the optical elements of proto-Lightspeed, and we describe the design in more detail below.

Light from the telescope first passes through a ZWO EFW filter wheel (Fig.~\ref{fig:cad_model}, component a), which may host up to seven filters of diameter 2~in. Filter options are discussed in Sec.~\ref{sec:filters}. An NS45B shutter from Uniblitz (component b) is affixed to this filter wheel, allowing for the capture of dark frames when closed. At the native telescope focus, a Zaber LSQ075A-E01T3A linear motorized stage (component c) can move a rectangular mask into the beam. This mask defines the field for single-shot linear polarimetry with proto-Lightspeed. This capability is not yet fully commissioned, but our strategy is discussed in Sec.~\ref{sec:polarimetry}.

A Canon EF 400~mm $f/4$ DO IS II USM lens (component d) then re-collimates the light. A stepper motor (component e) is connected to the focus ring of this lens by a belt and custom 3D-printed gears. This allows for internal focus control of the instrument. A fold mirror (component f) reflects the collimated beam toward the remaining instrument components. Narrow-band filters may be moved into the collimated beam via a Zaber LSQ150A-E01T3A stage (component g). At present, an $H_\alpha$ filter manufactured by Alluxa with diameter of 50~mm, center wavelength of 653.3~nm, and FWHM of 1.1~nm is mounted here but could be replaced by other narrow-band filters of interest. This stage will also host a superachromatic quarter wave plate (QWP) with diameter 2~in (SAQ-200-420/1100 from Meadowlark Optics) for polarimetry. Also for polarimetry (see Sec.~\ref{sec:polarimetry}), a second Zaber LSQ150A-E01T3A stage (component h) can move a wedged double Wollaston (WeDoWo) into the beam.

The collimated light is then re-imaged by a Canon RF 100--300~mm $f/2.8$ L IS USM lens (component i). A second stepper motor and gearing assembly (component j) is used to adjust the focal length of this lens and thereby control instrument zoom. Section~\ref{sec:reimaging} provides more details about the collimating and re-imaging lenses. An RF controller from Birger Engineering (component k) mates with the RF connector of the re-imaging lens, allowing for control of the focus and image stabilization mechanisms of the lens. We machined the face of this controller and the face of the ORCA-Quest~2 camera (component l) and fastened the two together, as shown in Fig.~\ref{fig:on_telescope}c, such that the camera's sensor is positioned at the back focal distance of the RF lens.

\subsection{Re-imaging Lenses}
\label{sec:reimaging}
We do not present a detailed optical prescription for proto-Lightspeed's COTS re-imaging lenses, as they have proprietary designs. However, the fundamental optical principle is straightforward: a backwards-facing telephoto lens with long focal length (400~mm) re-collimates light from the telescope focal plane, then a telephoto lens with shorter focal length (100--300~mm) yields a de-magnified image.


When the focal length of the re-imaging lens is minimized to 100~mm, we achieve the requisite de-magnification factor of 4 and a pixel scale of $0.050''$/pix. The focal length can be increased to yield smaller pixel scales, with a minimum of $0.017''$/pix, suitable for speckle interferometry. The lenses achieve re-imaging of a $1'$ diameter field from the native telescope focus with no worse than $30\%$ loss from vignetting. The full vignetted field extends to $\approx 1.5'$ in diameter. This corresponds to an image circle on the sensor with a diameter of $\approx 1800$~pix for maximum de-magnification and $\approx 5430$~pix for minimum de-magnification. Thus, at maximum de-magnification, the field utilizes $\approx 30\%$ of the sensor area; at minimum de-magnification, the full sensor area is used and covers a field of $0.65'\times 1.16'$.

The beam exiting the collimating lens has a size of $\approx 70$~mm. However, the beam reaches a pupil with diameter of 36~mm at a distance of $\approx 750$~mm from the collimating lens before diverging. The fold mirror, positioned 463~mm from the face of the collimating lens, minimizes the size required for optical components within the collimated beam while keeping the instrument compact. Polarization optics and narrow-band filters are positioned near the collimated beam pupil, $\approx 287$~mm from the fold mirror.

\subsection{Spectral Filtering}
\label{sec:filters}
The ZWO EFW filter wheel is populated with photometric filters ($u', g', r', i', z'$) and one filter for speckle interferometry centered at the OIII line of 500.7~nm (near the QE peak of the detector) with a full width at half maximum (FWHM) of 9~nm. All of these filters were manufactured by Baader Planetarium. The $u', g', r', i', z'$ filters have the same center wavelength and bandwidths as the original primed filter set of the Sloan Digital Sky Survey \citep[SDSS;][]{doi:2010}, but with slightly different spectral response (in particular, slightly higher throughput for shorter wavelengths in the $g'$ band). One filter position is left open.

\begin{figure}
    \centering
    \includegraphics[width=0.99\linewidth]{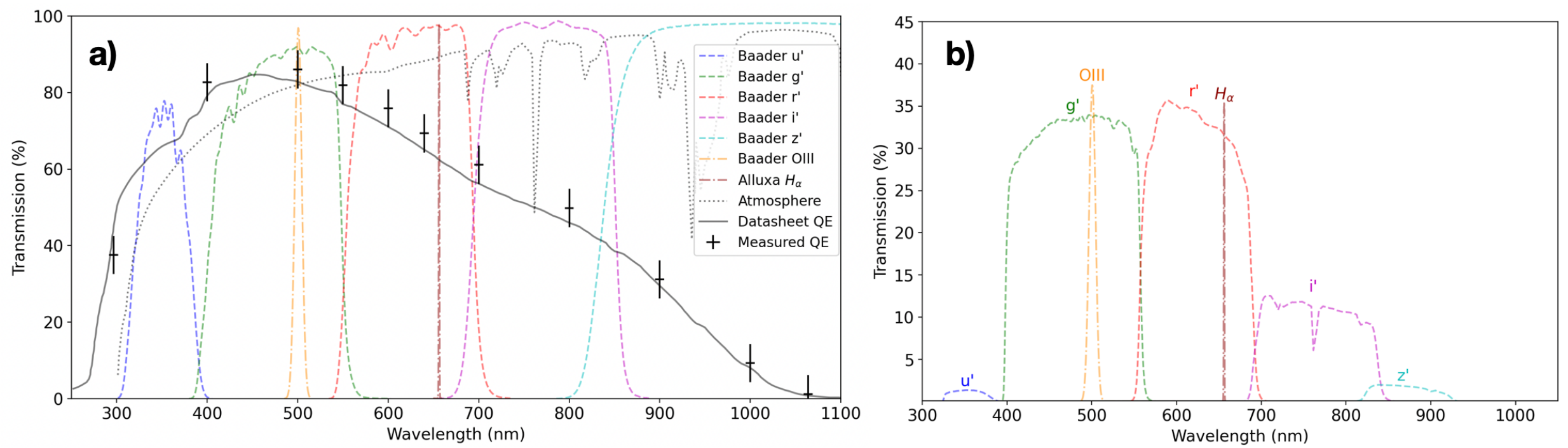}
    \caption{\textbf{a)} Transmission curves for filters available in proto-Lightspeed (dashed lines), the quantum efficiency of the ORCA-Quest 2 camera (crosses and solid line), and atmospheric transmission at unit airmass (dotted line). \textbf{b)} Total throughput for each bandpass. This includes the filter transmission, sensor quantum efficiency, losses due to the re-imaging optics, and atmospheric transmission at unit airmass. It does not include the transmission of the telescope. The curves for $u'$ and $z'$ are upper limits, with an assumed re-imaging throughput of 5\%.}
    \label{fig:throughputs}
\end{figure}

Figure~\ref{fig:throughputs}a shows the bandpasses of the proto-Lightspeed filter options, while Fig.~\ref{fig:throughputs}b shows the total throughput in the center of the field with each filter choice, including losses due to the re-imaging optics and fold mirror, atmospheric transmission at unit airmass, and the sensor's QE. We measured the throughput of the re-imaging optics in each band in the laboratory by forming a broadband point source (using a fiber-coupled laser-driven light source, model EQ-99X-FC from Energetiq) at the focus of the collimating lens, with the filter under test between this source and collimating lens. Using a Thorlabs S130C photodiode power sensor, we then measured the optical intensity in front of the collimating lens and at the image formed by the re-imaging lens. The throughput for the $i'$ channel is poor, due to the coatings and materials of the COTS re-imaging optics. For $u'$ and $z'$, we measured a throughput consistent with zero, as expected for the same reason. For these bands we report only upper limits corresponding to the noise level of the power sensor relative to the power measured in front of the lenses ($\approx5\%$ for both bands). Thus, the $u'$ and $z'$ filters should generally not be used.

\subsection{Polarization Optics}
\label{sec:polarimetry}
proto-Lightspeed is designed to allow for single-shot measurement of linear polarization and measurement of circular polarization. This polarimetry subsystem has not yet been integrated into instrument operation, but here we discuss its design. proto-Lightspeed will employ a wedged double Wollaston (WeDoWo) prism, proposed by \cite{oliva:1997} and used in multiple astronomical instruments, first by \cite{pernechele:2003}. Such a WeDoWo prism splits a masked region of the field into four images on the detector, corresponding to polarization angles of $0^\circ$, $90^\circ$, $45^\circ$, and $135^\circ$. proto-Lightspeed will use a custom-fabricated crystal quartz WeDoWo from the Karl Lambrecht Corporation, but its specifications are still being determined. Each Wollaston prism (one $90^\circ/0^\circ$ and one $45^\circ/135^\circ$ prism) will have a beam deviation between $1^\circ$ and $2^\circ$. The wedges bonded to these prisms will have angles between $3^\circ$ and $5^\circ$. When both the QWP and WeDoWo are positioned in the beam, the circular polarization can be probed. The field-defining mask for polarimetry has open dimensions of $20\,\mathrm{mm} \times 5\,\mathrm{mm}$, corresponding to $15'' \times 1'$.

\section{Mechanical and Electrical Design}
\label{sec:mech_design}

\begin{figure}[b!]
    \centering
    \includegraphics[width=0.95\linewidth]{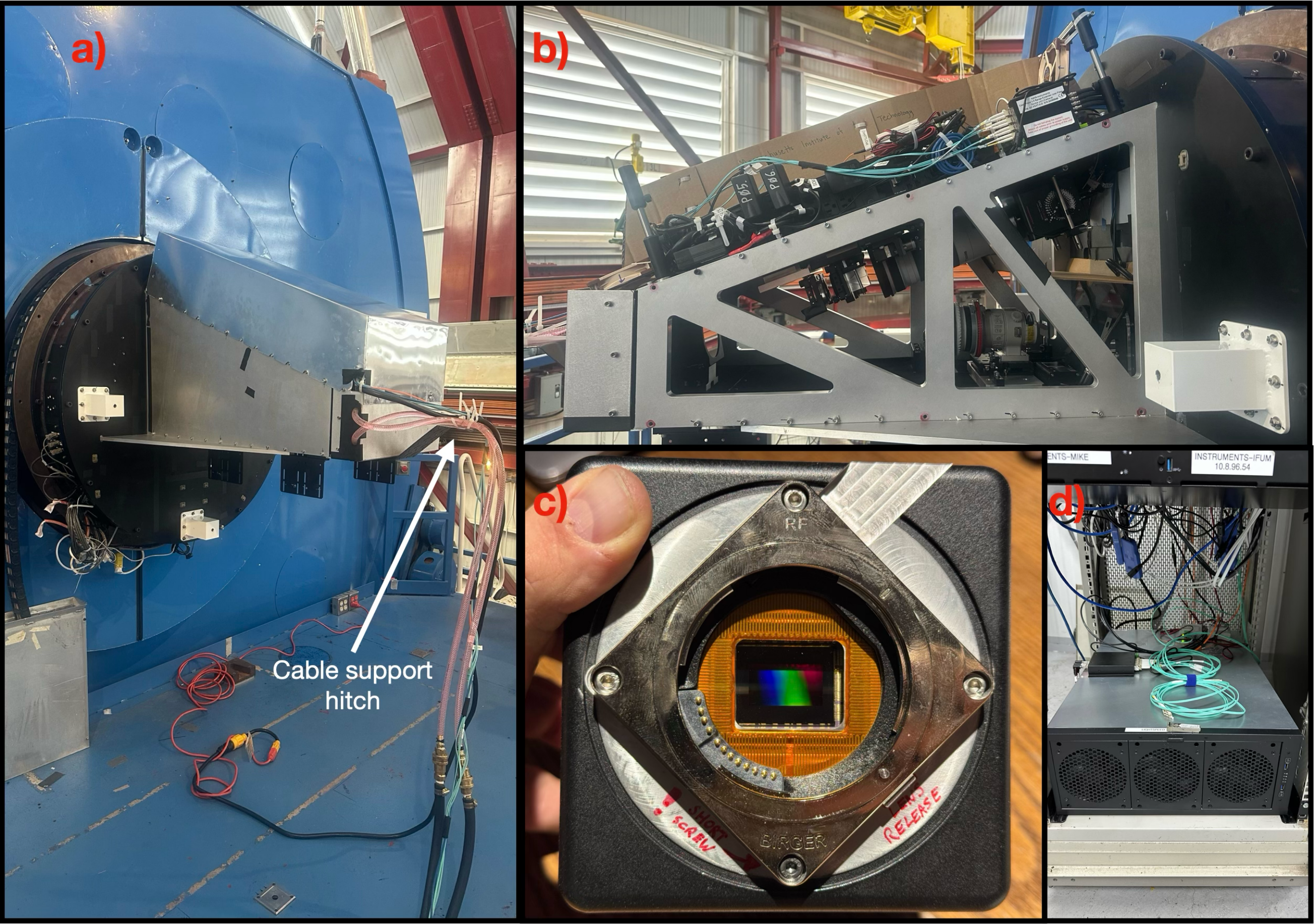}
    \caption{Overview of proto-Lightspeed components. \textbf{a)} proto-Lightspeed mounted at the NasE port of the Clay telescope. The fully assembled instrument extends $\approx$1~m from the port. Five LC duplex fibers, power and Ethernet cables, and glycol cooling lines are secured to an aluminum hitch at the optical axis to prevent stress during rotation. \textbf{b)} Interior view with cooling enclosure and side panels removed, showing the dual Thorlabs breadboards, optical components, and cardboard airflow baffles that direct cooling air across electronics and to the camera TECs. \textbf{c)} The ORCA-Quest 2 camera, with the Birger RF lens controller affixed. \textbf{d)} The control computer mounted in the Clay equipment room, housing the Active Silicon Firebird CoaXPress frame grabber and Meinberg TCR180PEX GPS timing card.}
    \label{fig:on_telescope}
\end{figure}

The optical components are mounted on two solid aluminum breadboards from Thorlabs of size $18\,\mathrm{in}\times 36\,\mathrm{in} \times 0.5\,\mathrm{in}$. These breadboards were modified for bolting to solid aluminum support structures. These support structures are secured to a solid aluminum rotator plate for direct mounting to the NasE port. This plate was previously manufactured to mount the prototype of the Large Lenslet Array Magellan Spectrograph (LLAMAS) on the same port \citep{furesz:2020}. This basic support structure is shown in Fig.~\ref{fig:on_telescope}b. A cart also manufactured for the LLAMAS prototype holds the instrument when not in use and allows for convenient installation on the telescope. Easily removable sheet aluminum panels enclose the sides of the instrument. All gaps and breadboard holes are covered with black felt adhesive.

The top side of the upper breadboard hosts most of proto-Lightspeed's electronic components. A Kaya Instruments KY-FEXT-II-D converts the four CoaXPress outputs of the ORCA-Quest 2 camera to four LC duplex fibers. Two Zaber X-MCC4 controllers drive the two stepper motors and three linear stages. A LabJack T4 sends transistor-transitor logic (TTL) signals to a Uniblitz ED12DSS controller to open and close the shutter. A Raspberry Pi single-board computer connects via USB to the Birger RF lens controller and the ZWO filter wheel. Fiber converters from Meinberg USA (one CON/FO/TTL-2/HS and one CON/TTL/FO/HS) connect to two timing ports on the ORCA-Quest 2 camera, yielding a single LC duplex fiber. A 16-plug CyberPower PDU41003 power distribution unit (PDU) powers all components. The PDU, LabJack, X-MCC4s, and Raspberry Pi are connected to a Maple Systems MS1-M08G managed ethernet switch.

A cooling system channels cold air across the electronic components and supplies fresh air to the fans and thermo-electric coolers (TECs) of the ORCA-Quest 2 camera. Simple cardboard pieces define airflow channels over the electric components and cordon the air around the camera from the rest of the instrument interior, as shown in Fig.~\ref{fig:on_telescope}b. There is a spacing of $\approx 1$~in between the top breadboard and the rotator plate, allowing efficient air supply to the camera. At the end of the instrument, a heat exchanger cools the circulating air with glycol supplied by the observatory. Air is pulled through the air channels and past the heat exchanger by four Noctua NF-A12x25 PWM fans. With this system, the ORCA-Quest 2 camera maintains its nominal operating temperature of $-20^\circ$C even when reading at its maximum frame rate for extended periods. For all observations in a five-night commissioning campaign in December 2025 (summer for the southern hemisphere), the sensor temperature remained steady at $-20^\circ$C.

An Ethernet cable, power cable, five LC duplex fibers, and the glycol cooling lines exit the instrument and are firmly affixed at the optical axis on an aluminum ``hitch," preventing stress during rotation, as shown in Fig.~\ref{fig:on_telescope}a. We use existing fiber connections on the NasE platform to pass data and timing information to and from the camera. The instrument extends $\approx 1$~m from the NasE port, with width and height (excluding the rotator plate) of 48~cm and 75~cm, respectively. Including the interface plate, the total mass of the instrument is $\approx 160$~kg.

The instrument control computer is mounted in the Clay telescope equipment room, as shown in Fig.~\ref{fig:on_telescope}d. Affixed on the computer, a Kaya Instruments KY-FEXT-II-D and two more Meinberg USA fiber converters receive LC duplex fibers from the telescope and connect to the control computer. In two PCI Express slots, the computer hosts an Active Silicon Firebird CoaXPress frame grabber card and a Meinberg USA TCR180PEX card. The former communicates with the ORCA-Quest 2 camera. The latter receives a GPS-synchronized IRIG timing signal from the Clay TCS computer via a BNC connection and provides the two timing connections to the camera. One line provides a GPS-synchronized pulse every second, used to trigger camera exposures. The other receives and time-tags signals sent from the camera whenever an exposure is completed.

\section{Detector}
\label{sec:detector}
The Hamamatsu ORCA-Quest 2 camera (product number C15550-22UP, often referred to as a ``qCMOS" camera) hosts the HWK4123 sensor developed by Fairchild Imaging, which has $4096\times 2304$ square pixels of size 4.6~µm. The camera has a specified full well capacity (FWC) of $7000\,\mathrm{e}^-$, dark current of $0.016\,\mathrm{e}^-$/pix/s at $-20^\circ$C, and peak QE of 85\% at 460~nm \citep{Hamamatsu_C15550_Catalog}. The HWK4123 sensor was first presented in \cite{cho:2023} and has been studied in the previous-generation Hamamatsu ORCA-Quest 1 camera by \cite{Khandelwal_2024} and \cite{gallagher:2024}.

proto-Lightspeed may use either of the ORCA-Quest 2 camera's two area readout modes, called standard scan and ultra-quiet scan. These modes provide 16-bit imaging with a specified root mean square (RMS) read noise of $0.43\,\mathrm{e}^-$ and $0.3\,\mathrm{e}^-$, respectively. The standard mode allows for faster imaging. Table~\ref{tab:readout_modes} shows the maximum readout rate for these two modes with different subarray settings. The maximum frame rate depends only on the number of rows, not columns. For proto-Lightspeed, we adopt a $32\times 1200$ pixel window, corresponding to $1.6''\times 1'$, as the smallest window size. Such a window fits $\approx 3$ times the FWHM of typical seeing and, when the rotator angle can be freely chosen, allows for the inclusion of at least one comparison star. As proto-Lightspeed's $1'$ field fits within $\approx 1200 \times 1200$ pixels, this full field can be imaged at 200~Hz in the standard mode and 48~Hz in the ultra-quiet mode.

\begin{table}[t!]
    \centering
    \begin{tabular}{c|c|c|c|c|c}
        Camera & Measured & \multicolumn{4}{c}{Maximum Frame Rate (Hz)} \\
        Readout & Read Noise$^*$ & Full Sensor & $1'$ Full Field & $1.6''\times 1'$ & Absolute Maximum \\
        Mode & (e$^-$ RMS) & $2304 \times 4096$~pix & $1200\times 1200$~pix & $32 \times 1200$~pix & $4 \times 4096$~pix \\
        \hline
         Standard & 0.41 & 120 & 200 & 6600 & 19800 \\
         Ultra-quiet & 0.29 & 25 & 48 & 1400 & 4200
    \end{tabular}
    \caption{Read noise and maximum frame rates for proto-Lightspeed's two readout modes. These two modes have nearly identical conversion gain, full well capacity ($\approx 7000\,\mathrm{e}^-$), and dark current. Frame rate values for other subarray sizes can be found at \citet{Hamamatsu_C15550_Catalog}. $^*$See Sec.~\ref{sec:read_noise} for discussion on measured read noise values.}
    \label{tab:readout_modes}
\end{table}

The standard and ultra-quiet modes each combine readouts from high ($\approx 32\times$) and low ($\approx 1\times$) gain paths in the analog signal chain. The high gain path enables ultra-low read noise but saturates for signals larger than $\sim 220\,\mathrm{e}^-$. Above this threshold, the camera uses the low gain path, which saturates at much larger signal values ($\sim 7000\,\mathrm{e}^-$). Signals output by the high gain path are scaled down by a gain correction factor to prevent a discontinuity at the switching threshold \citep{gallagher:2024,cho:2023,Hamamatsu_C15550_technical_note}. The ORCA-Quest 2 camera has an additional readout option, called ``photon number resolving," which uses the same scan speed as the ultra-quiet mode but employs only the high gain signal path. It therefore has a very limited dynamic range ($\sim 200\,\mathrm{e}^-$). This option rounds the digital output in analog-to-digital units (ADU) to the expected number of electrons collected. The inherent read noise prior to rounding should be the same as the ultra-quiet mode, but for individual pixels with read noise below $\approx 0.3\,\mathrm{e}^-$, rounding to the closest integer electron can yield a lower effective read noise. However, this method is inefficient in pixels where read noise is above $0.3\,\mathrm{e}^-$, when rounding can actually worsen the effective read noise. Furthermore, for rapid readout of faint sources where most pixels receive no photons, an ideal analysis would cut read noise more harshly to avoid drowning out the signal. The optimal cut depends on the source flux. For proto-Lightspeed, we therefore elected not to use the ORCA-Quest 2 photon number resolving option. Instead, we are developing a method which fits the source flux to the measured raw ADU distribution in each pixel to derive the most precise possible estimate of the number of photoelectrons. This method delivers an improvement over the raw signal and over rounding for read-noise dominated observations (see section \ref{sec:science_highlights}). A follow-up paper will describe this method.

We conducted our own characterization of the ORCA-Quest 2 camera, measuring its linearity, conversion gain, full well capacity (FWC), read noise, dark current, and QE. For all of these measurements, we used the uniform illumination optical system described in \cite{layden:2025}. Table~\ref{tab:detector_specs} summarizes the results of these measurements.

\begin{table}[h!]
    \centering
    \begin{tabular}{c|c|c}
        Parameter & Manufacturer specification & Measured value \\
        \hline
         Linearity error & 0.5\% (5--95\% saturation)$^*$ & See Sec.~\ref{sec:nonlinearity}\\
         Full well capacity (e$^-$) & 7000 & 7002 \\
         Read noise (e$^-$, RMS) & 0.30 & 0.29 \\
         Dark current (e$^-$/pix/s, mean, $-20^\circ$~C) & 0.016 & $0.005\pm 0.001$ \\
         Quantum efficiency ($\lambda=500$~nm) & 83\% & $86\pm5$\%$^\dag$ \\
    \end{tabular}
    \caption{Specifications and measured values for important parameters of the ORCA-Quest~2 camera. All values are specified for the ultra-quiet scan mode. Manufacturer specifications retrieved from \citet{Hamamatsu_C15550_Catalog}. $^*$The manufacturer follows the EMVA 1288 standard procedure for measuring linearity error, which only probes linearity between 5--95\% saturation \citep{emva1288}. This standard is not sufficient for astronomical applications involving faint sources. $^\dag$The internal quantum efficiency is degraded at low signal levels, causing nonlinearity. See Sec.~\ref{sec:nonlinearity_interp} for extended discussion.}
    \label{tab:detector_specs}
\end{table}

\subsection{Nonlinearity measurement and calibration}
\label{sec:nonlinearity}
Early in sensor testing, we noticed that the sensor exhibited significantly nonlinear response at very low exposure levels. This nonlinearity can significantly degrade photometric accuracy for faint sources, so a calibration must be developed to apply to raw science images. Such nonlinearity in the ORCA-Quest 1 camera was also observed by \cite{lucas:2024}, who mention but do not describe an approximating function for correcting it. They also do not discuss the cause of this nonlinearity or how the application of such a correction can affect the noise properties of science images. We previously observed similar but more severely nonlinear response in the Teledyne COSMOS CMOS image sensor, likely resulting from incomplete charge transfer between the photodiode and sense node of each pixel. For that camera, we developed a calibration strategy for recovering linear response and noted the importance of appropriately modeling noise sources \citep{layden:2025}. We aim to do the same here for the ORCA-Quest 2 camera. This nonlinearity calibration also accounts for any discontinuity when the camera switches from the high gain path to the low gain path, which would result from imperfect gain correction factors. Such a discontinuity was previously observed from a switching in gain paths in the Marana CMOS camera \citep{apergis:2025}.

\subsubsection{Measurement procedure}
To collect the data necessary to measure the nonlinear response of the camera and develop a nonlinearity calibration, we followed a similar procedure as in \cite{layden:2025} for the ultra-quiet readout mode. We acquired a sequence of images at increasing exposure times, logarithmically spaced from the minimum sensor exposure time to an exposure time yielding saturation, with a filter centered at 640~nm. The sensor temperature was maintained at $-20^\circ$C. At each exposure time, we acquired a stack of 25 frames and calculated a mean frame from this stack. We also acquired a stack of 50 bias frames, with the exposure time set to a minimum (34~µs) and the lens cap on. We then calculated the mean bias frame, which we subtracted from the mean frame at each exposure time.

\begin{figure}
    \centering
    \includegraphics[width=0.98\linewidth]{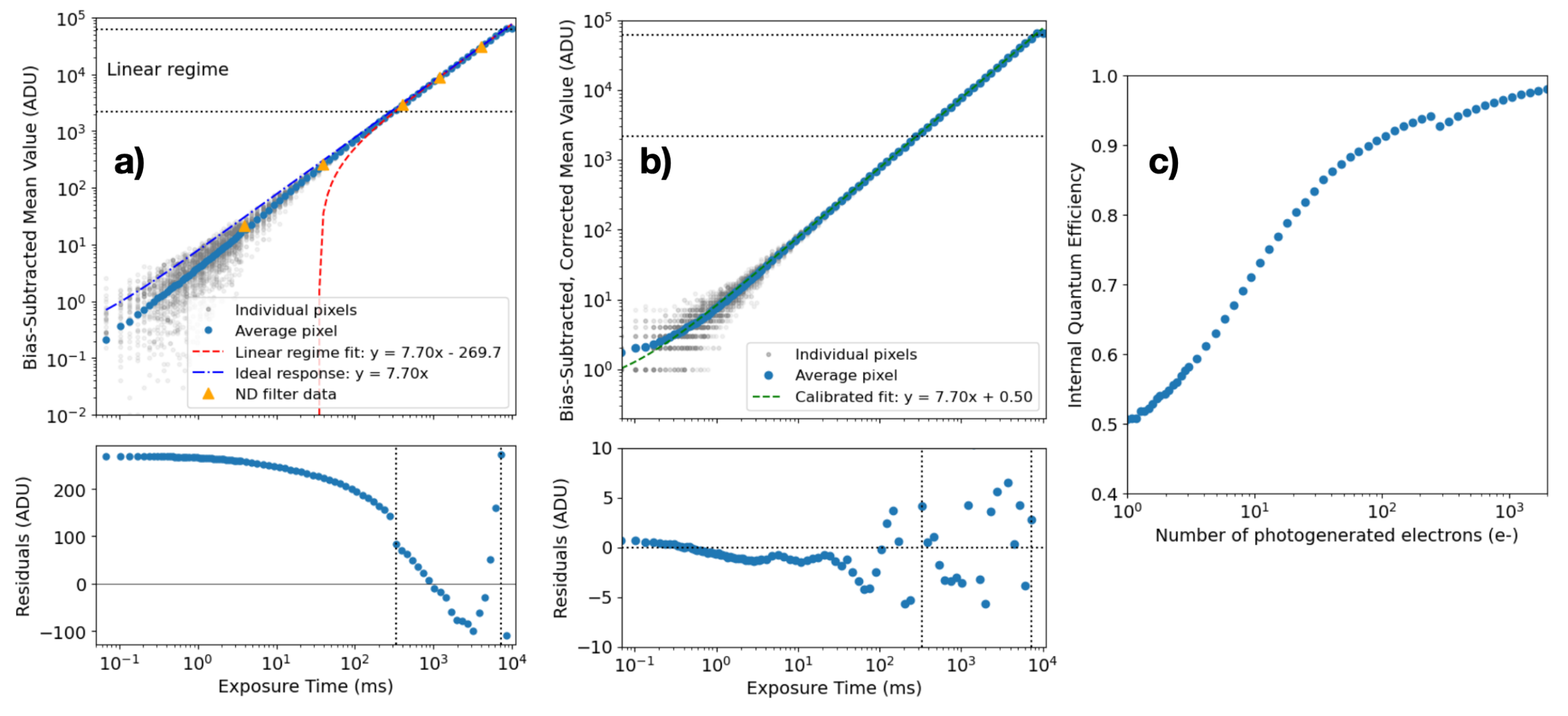}
    \caption{Nonlinearity and internal QE of the ORCA-Quest 2 camera. \textbf{a)} \textit{Top panel:} The raw sensor response to increasing exposure times from a stable uniform light source, averaged across the sensor (blue points), shows significant nonlinearity at low signal levels. Individual pixels (gray points) all have slightly different response. The red dashed line shows a linear fit to the regime where the sensor response is close to linear (2400\,ADU--62000\,ADU, indicated by dashed black lines). The blue dash-dotted line shows the ideal linear response, with the same slope but zero y-intercept. Orange triangles show the average response to 4~s exposures with varying neutral density filters, confirming the nonlinearity is not a timing artifact. \textit{Bottom panel:} Residuals from the linear fit clearly show poor agreement at low signals and a discontinuity at $\approx$2400~ADU where the camera switches gain paths. \textbf{b)} \textit{Top panel:} After applying the per-pixel nonlinearity calibration, the sensor response is linear across the full dynamic range. The green dashed line shows the fit to this corrected data. Residual spread in individual pixel response (gray points) is caused by read noise and shot noise. \textit{Bottom panel:} Residuals remain below 1 ADU at the lowest signal levels. \textbf{c)} Signal-dependent detection efficiency, showing the fraction of photogenerated electrons that successfully reach the sense node. At low signals, incomplete charge transfer causes $\sim$50\% of photoelectrons to be trapped; this efficiency improves asymptotically to nearly 100\% at higher signal levels.  The calibration compensates for this effect. The slight discontinuity in this curve is an artifact from the gain shift.}
    \label{fig:nonlinearity}
\end{figure}

The blue points of Fig.~\ref{fig:nonlinearity}a show the bias-subtracted mean value (averaged over all pixels) at each exposure time. The gray points show the response for 50 individual pixels, demonstrating slightly different nonlinear behavior in each. We observed a discontinuity in the average pixel response at 2400~ADU, likely resulting from the high gain path saturating and the camera switching to the low gain path. We fit a line (dashed red curve) to the data above this discontinuity and up to 62000~ADU, beyond which saturation can affect the mean value. The best fit line to this regime, where the sensor response is nearly linear, has a y-intercept well below zero. As discussed in Sec.~\ref{sec:nonlinearity_interp}, this is a sign of incomplete charge transfer. The bottom panel of Fig.~\ref{fig:nonlinearity}a shows the residuals between the measured response and this fit to the nearly linear regime, clearly demonstrating the aforementioned discontinuity and the poor agreement for low signals. Even in the regime to which a line was fit, a low level of nonlinearity is still present, as indicated by the structure of the residuals here.

To confirm that this nonlinearity was not caused by inaccurate camera timing at short exposures, we repeated the measurement at a fixed exposure time of 4~s using Thorlabs neutral density (ND) filters (optical densities 0.5, 1, 2, and 3) to yield a range of low signal levels. The orange triangles in Fig.~\ref{fig:nonlinearity}a show the mean values measured with these ND filters, plotted at effective exposure times found by scaling 4~s by the nominal optical density at 640~nm. The agreement with the short-exposure data (blue points) confirms that the nonlinearity is not an exposure timing artifact.

\subsubsection{Calibration procedure}
The blue dash-dotted line in Fig.~\ref{fig:nonlinearity}a shows the ideal response of the sensor, with the same slope as the red line and a y-intercept of zero. The goal of the nonlinearity calibration is to map raw signal values onto this ideal response. This calibration must be constructed for each individual pixel, as each has a slightly different response curve.

We adopt a per-pixel nonlinearity calibration procedure similar to that of \citet{layden:2025}, which we describe briefly below. The operating range of the pixel was split into three regimes ($x\leq 17$~ADU, $12\,\mathrm{ADU}<x\leq 2250\,\mathrm{ADU}$, and $x>2200$~ADU for signal level $x$). A polynomial fit for each regime was constructed to map the measured data onto the ideal response. The overlap between the three regimes improves the smoothness of the calibration. The orders of the polynomials fit to each regime were 3, 8, and 6, respectively, and were empirically optimized to minimize residuals while avoiding overfitting. When the calibration is applied, for raw values in $12\,\mathrm{ADU} < x < 17\,\mathrm{ADU}$, a linear interpolation between the values calculated by the low and middle polynomial fits is applied; a similar interpolation is applied between the middle and high polynomial fits for raw values in $2200\,\mathrm{ADU} < x < 2250\,\mathrm{ADU}$. After a fixed bias level of 200~ADU is added to calibrated values to prevent clipping of negative values, the calibrated image is saved with the same data format as raw frames (using unsigned 16-bit integers). While these polynomial fits are empirical rather than physically motivated, they are computationally efficient and flexible enough to achieve excellent calibration for each individual pixel across the full dynamic range. Figure~\ref{fig:nonlinearity}b shows the average signal value reported at each exposure time after the calibration has been applied, demonstrating excellent linearity even at the lowest signal levels, where residuals remain below 1~ADU.

We have not yet developed a nonlinearity calibration for the standard (higher speed) readout mode, as hardware issues delayed its commissioning until December 2025. Initial testing suggests the nonlinear response is nearly identical between modes, as was observed in the COSMOS camera \citep{layden:2025}. We are validating whether the ultra-quiet mode calibration can be applied to this mode with minor adjustments. Until this is complete, we recommend using the ultra-quiet mode for any observations that do not demand readout speeds exceeding its capabilities (48~Hz full field/1400~Hz windowed).


The bias levels of individual pixels can depend very weakly (by 1--2 ADU) on the subarray selected for imaging. Thus, when high precision is required, an average bias frame should be taken using the same subarray as the science images. The pixel-wise nonlinear response should not change with subarray selection.

\subsubsection{Interpretation of the nonlinearity}
\label{sec:nonlinearity_interp}
As discussed in \cite{layden:2025}, nonlinear response at low light levels can be explained by incomplete charge transfer in each pixel. This incomplete charge transfer could result from traps along the transfer channel or from a local extremum in the electric potential along the path of charge transfer, referred to as a barrier or pocket \citep{stefanov:2022}. Due to manufacturing variance, each pixel can have a different level of incomplete charge transfer, necessitating a per-pixel calibration. The nonlinear response for each pixel is stable over time but can vary slightly with temperature or pixel voltage settings, as these affect the probability for an electron to escape from a trap or pocket.

As a result of incomplete charge transfer, the QE of the sensor is effectively degraded and dependent on the number of photoelectrons that are generated in a pixel. For a small charge packet collected in the photodiode, a large fraction of this charge packet may be caught in traps or a pocket during charge transfer.  For a large charge packet, a small fraction of the charge packet will saturate the traps or pocket, yielding only a small QE degradation. The negative y-intercept of the best fit line to the sensor response (red dashed curve in Fig.~\ref{fig:nonlinearity}a) represents the saturation level of these traps, which we found to be $\approx 30\,\mathrm{e}^-$ (using the conversion gain measured in Sec.~\ref{sec:ptc}).

The effect of the nonlinearity calibration is to scale up small signal values, making the QE constant. Therefore, the camera's internal QE for a given signal level can be calculated as the inverse of the multiple by which such a signal level is scaled up by the calibration. Figure~\ref{fig:nonlinearity}c shows this signal-dependent detection efficiency for low signal levels, normalized to the QE measured at high signal levels.

Although the calibration scales up small signal levels to recover an accurate measure of the photon flux, it also scales up the noise associated with the small number of electrons that actually reach the sense node. Because the lowest signal values must be scaled by a factor of $\approx 2$ (see Fig.~\ref{fig:nonlinearity}c), bias level fluctuations are also scaled up by this factor in faint calibrated images (and shot noise by the square root of this factor). Noise calculations involving calibrated data must account for this scaling. This performance at the lowest count rates is equivalent in SNR to a perfectly linear sensor with 50\% worse QE, though the performance improves for higher count rates. For ultra-faint or ultra-fast observations, the expected number of photons per frame per pixel may be no more than a few. If this is true, one might consider ignoring the nonlinearity calibration and simply treating the sensor as having $\approx 50\%$ worse QE. We in particular recommend this strategy for observations taken with the standard readout mode, for which a nonlinearity calibration is still under development.

\subsubsection{Effect of temperature on nonlinearity}
\label{sec:nonlinearity_temp}
Assuming traps or pockets cause the sensor's nonlinearity, this nonlinearity should be less severe at higher sensor temperatures (where electrons have a greater probability of escaping traps or pockets and reaching the sense node). We tested this hypothesis by collecting very short (1\,ms) exposures using our stable light source at temperatures between $-20^\circ$C and $35^\circ$C, in increments of $5^\circ$C. Unfortunately, to our knowledge, there is no way to specify a temperature set point for the ORCA-Quest~2 camera. Assuming adequate air flow (and no liquid cooling), with the TEC and fans enabled, the sensor will maintain a temperature of $-20^\circ$C. To vary the temperature, we first cooled the sensor to $-20^\circ$C, then disabled the TEC and fans and collected exposures as the sensor heated. This heating was slow enough that the temperature was close to constant during each exposure (varying by $1^\circ$C every few seconds). As discussed in Sec.~\ref{sec:dark_current}, we also collected a dark frame with exposure time 2\,s at each temperature. While the dark current increased by a factor of $\sim100$ as the temperature rose from $-20^\circ$C to $35^\circ$C, it remained a negligible contribution to the 1\,ms exposures.

With the 1\,ms exposures, we probed the internal QE at the faintest levels of exposure for each temperature. Using the ``ideal response" equation in Fig.~\ref{fig:nonlinearity}a, if the sensor had no charge transfer losses, then at an exposure time of 1\,ms, the mean bias-subtracted signal value would be 7.7\,ADU. The mean value at $-20^\circ$C and 1\,ms was 4.1\,ADU, corresponding to an internal QE of 53\%---this is the y-intercept of Fig.~\ref{fig:nonlinearity}c. The mean value of the 1\,ms frames increased roughly linearly from 4.1\,ADU to 5.7\,ADU over the $-20^\circ$C to $35^\circ$C range. This corresponds to a minimum internal QE of 74\% when the temperature is $35^\circ$C---a significant improvement compared to $-20^\circ$C. Thus, temperature can indeed significantly affect the response of sensors with low-signal nonlinearity. We operate proto-Lightspeed at $-20^\circ$C because of our requirements for temperature stability and low dark current, so we have not yet probed the full nonlinear response curve at different temperatures, but the effect warrants further study.

\begin{figure}
    \centering
    \includegraphics[width=0.8\linewidth]{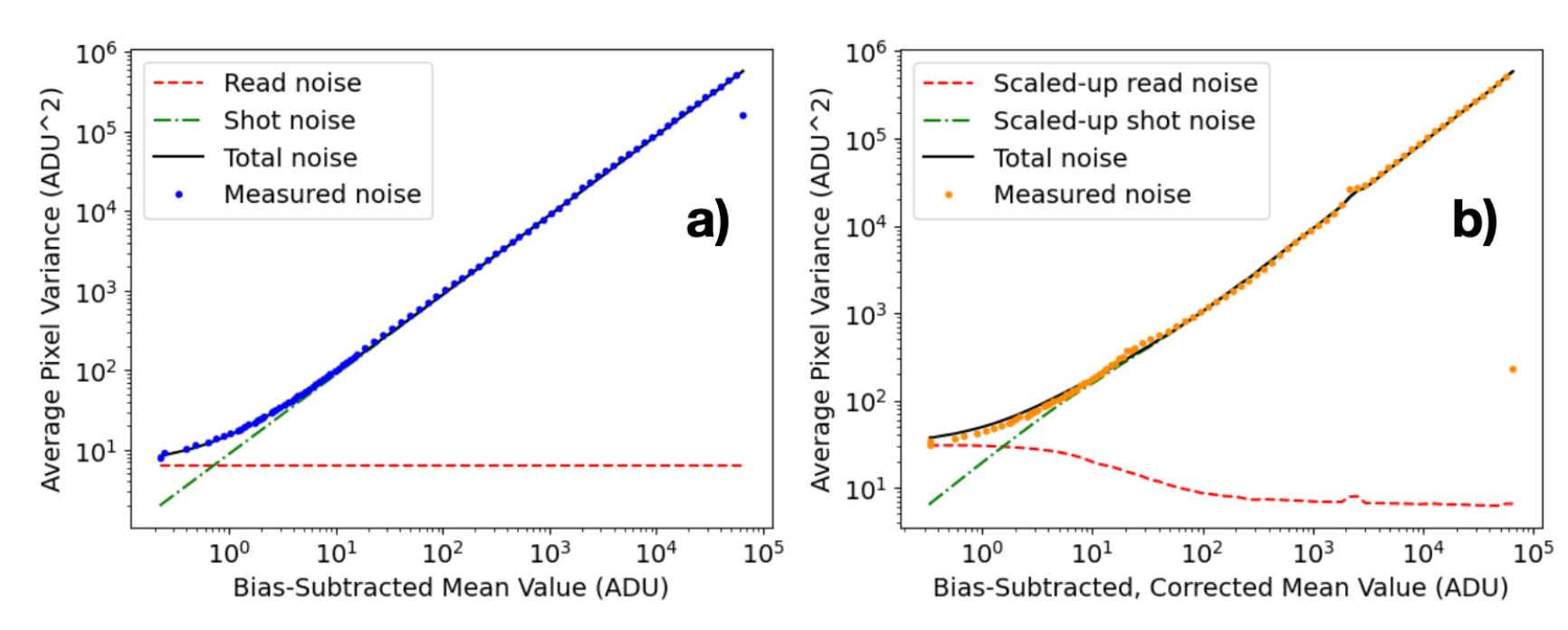}
    \caption{\textbf{a)} PTC for raw frames at increasing levels of exposure, with contributions from read noise and from Poissonian shot noise. \textbf{b)} Photon transfer curve (PTC) for images that have been corrected for linearity. Here the contributions from read noise and from Poissonian shot noise have been scaled to appropriately account for the effect of the linearity calibration. The bumps around 2500~ADU in \textbf{a)} and \textbf{b)} result from the transition between high and low gain paths.}
    \label{fig:ptc}
\end{figure}

\subsection{Photon transfer curve; conversion gain; full well capacity}
\label{sec:ptc}

To measure the conversion gain of the sensor and confirm our understanding of noise sources, we created a photon transfer curve \citep[PTC;][]{janesick2007photon} using the same dataset that was used to measure the sensor nonlinearity. For every exposure time, at which 25 raw frames were collected, we measured the mean and variance of each pixel's signal value. We then plotted the average pixel variance value against the average pixel mean value, as shown in Fig.~\ref{fig:ptc}a. We fit the region of this curve between 25\% and 75\% of saturation (where we observed the sensor response to be linear) to a line. The slope of this line gives the conversion gain, which we measured as $8.9\,\mathrm{ADU/e}^-$. We also applied the nonlinearity calibration to all frames and created a PTC using these frames, shown in Fig.~\ref{fig:ptc}b. The slope of this PTC from calibrated images between 25\% and 75\% of saturation was also $8.9\,\mathrm{ADU/e}^-$, as expected. In Fig.~\ref{fig:ptc}a and Fig.~\ref{fig:ptc}b, bumps are present near 2200\,ADU and 2500\,ADU, respectively. These bumps correspond to to the slight change in conversion gain (and, likely, in read noise) when the sensor readout switches to exclusively use the low gain path, as discussed in Sec.~\ref{sec:detector}. This bump is more prominent in the PTC for calibrated images, and a second small bump is observed in this PTC near 20~ADU. Both of these bumps exist near the transitions between polynomial fits of the nonlinearity calibration and likely result from those transitions not being perfectly smooth.

Comparing Fig.~\ref{fig:ptc}b to Fig.~\ref{fig:ptc}a clearly demonstrates how, in calibrated images, the noise components do not follow traditional scaling relations. To account for the sensor effectively having a signal-dependent degradation in QE, low signal levels (as well as their associated read noise and photon shot noise, shown in red and green, respectively, in Fig.~\ref{fig:ptc}b) are scaled up. This is important to consider when estimating the SNR of observations, particularly for faint objects. The proto-Lightspeed exposure time calculator accurately models this noise, as discussed in Sec.~\ref{sec:etc}.

We have not yet created a PTC for the standard readout mode. We assume that the standard and ultra-quiet modes have the same conversion gain to within $5\%$, as per Hamamatsu's specifications of the camera unit.

The full well capacity (FWC) is defined as the maximum number of photoelectrons that each pixel can hold. Since bit saturation often occurs before the physical full well capacity is reached, in these instances the saturation capacity (the number of electrons yielding the maximum digital output of the sensor) is typically reported as the FWC. This is the case for the ORCA-Quest 2. Because the mean value of flat illuminations near saturation will be skewed by the saturation, it is difficult to develop a nonlinearity calibration that works reliably for raw signal values very near saturation. For this reason, we set a maximum bias-subtracted value for calibration to be applied at 62,070~ADU$_{raw}$. For any raw signal values greater than this, the calibration returns a saturated signal value of $2^{16}-1=65535$~ADU. The raw upper limit, when calibrated, yields a value of 62,326 ADU (almost exactly $30\,\mathrm{e}^-\times 8.9\,\mathrm{ADU/e}^-$ greater than the raw value, as expected for the trap capacity measured in Sec.~\ref{sec:nonlinearity_interp}). Dividing this value by the conversion gain yields a FWC of $7003\,\mathrm{e}^-$, in agreement with Hamamatsu's specifications.

\subsection{Read noise}
\label{sec:read_noise}
We measured the read noise of the camera using 300 frames taken at minimum exposure time (34~µs) in a dark enclosure. We measured the standard deviation across these frames for each pixel before and after applying the nonlinearity calibration. We divided these values by the conversion gain to derive the read noise of each pixel in e$^-$. For the ultra-quiet mode, we found the RMS pixel read noise value for raw images to be $0.29\,\mathrm{e}^-$, in agreement with manufacturer specifications. After the calibration is applied, the RMS pixel read noise is $0.58\,\mathrm{e}^-$. This read noise in calibrated images reflects an ``effective" read noise that results from a 0.29~e$^-$ sensor operating at effectively 50\% QE for low light levels. That is, the detector readout \textit{does} deliver an RMS read noise below 0.3~e$^-$ and therefore \textit{can} still nearly resolve individual photoelectrons reaching the sense node. A fraction of photoelectrons (an especially large fraction at low signal levels) simply don't reach the sense node, so the read noise when referred to this number of generated photoelectrons is larger: $0.58\,\mathrm{e}^-$. As discussed in Sec.~\ref{sec:detector}, we are developing a data processing step to make use of the separation of peaks in raw signal levels corresponding to discrete numbers of photoelectrons. This step could significantly reduce the effective read noise in raw and calibrated images. We measured the raw RMS read noise using the standard mode to be $0.41\,\mathrm{e}^-$, using 300 raw bias frames. As this read noise is noticeably higher than the ultra-quiet mode (effectively preventing counting of individual photoelectrons), we encourage the use of the ultra-quiet mode whenever possible.

\subsection{Dark current}
\label{sec:dark_current}

We measured the dark current of the detector by taking a series of proto-Lightspeed exposures, with the shutter closed and all dome lights turned off. We used the ultra-quiet mode, with exposure times of 1, 10, and 30 minutes, at the stable operating temperature of $-20^\circ$C. We applied the calibration to these images to accurately measure the number of thermally generated photoelectrons. We measured the dark current at $-20^\circ$C to be $0.005\pm 0.001\,\mathrm{e}^-$/pix/s, or $18\pm 4\,\mathrm{e}^-$/pix/hr. While this is higher than the value one would infer before applying the nonlinearity correction, it is still lower than the manufacturer specification. Even on the darkest nights, this rate is an order of magnitude lower than the sky background for the $g'$, $r'$, and $i'$ filters and lower than the sky background even for the narrow-band $H_\alpha$ filter (Sec.~\ref{sec:etc} discusses how the sky background is modeled).

\begin{figure}
    \centering
    \includegraphics[width=0.98\linewidth]{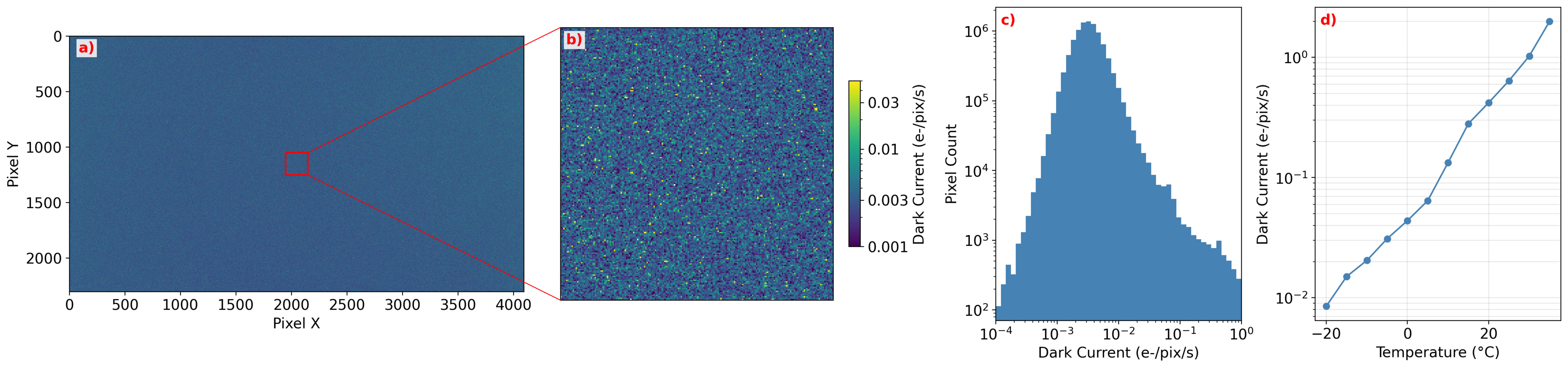}
    \caption{Dark current properties of the ORCA-Quest~2 camera. \textbf{a)} The dark current measured in each pixel at $-20^\circ$C, showing no clear patterns or regions of high glow. \textbf{b)} A zoom-in on a 200$\times$200 pixel subregion, showing the uniform distribution of hot pixels. \textbf{c)} The distribution of pixel dark current values at $-20^\circ$C, showing a somewhat extended tail of hot pixels. \textbf{d)} Average pixel dark current vs. temperature. These values should be taken as overestimates, as temperature was not well controlled.}
    \label{fig:dc}
\end{figure}

Figure~\ref{fig:dc}a shows a map of the dark current in the detector pixels, with Fig.~\ref{fig:dc}b zooming into a 200$\times$200 pixel subregion. These images show no clear patterns in the dark current and no evidence of amplifier glow; hot pixels appear uniformly distributed across the sensor. Figure~\ref{fig:dc}c shows the distribution of pixel dark current values. While the distribution peaks below $0.005\,\mathrm{e}^-$/pix/s, there is an extended tail of hot pixels.

We also roughly probed the temperature dependence of the dark current by taking 2~s exposures at temperatures between $-20^\circ$C and $35^\circ$C, in increments of $5^\circ$C. Because the temperature was rising during exposures (Sec~\ref{sec:nonlinearity_temp} discusses how we relied on gradual uncontrolled camera heating to reach each temperature), these values should be taken as overestimates. To remove the effect of temperature-dependent internal QE, we divided the measured dark current at each temperature by the minimum internal QE measured for Sec.~\ref{sec:nonlinearity_temp}. The results are shown in Fig.~\ref{fig:dc}d. Because we did not construct a complete nonlinearity calibration at each temperature and because the temperature was not well controlled, these results should be taken as estimates.

\subsection{Quantum efficiency}
\label{sec:qe}

We measured the QE at eleven wavelengths from 297--1064~nm to confirm the reliability of the higher resolution QE profile provided by Hamamatsu (\cite{Hamamatsu_C15550_Catalog}). To do so, we used the optical apparatus consisting of a stable light source, spectral filters, and calibrated photodiode described in \cite{layden:2025}. We measured the QE of a second ORCA-Quest 2 camera that we had access to, as the proto-Lightspeed camera had already been integrated into the instrument. Figure~\ref{fig:throughputs}a shows our measured QE values alongside values specified in the Hamamatsu datasheet. The measured QE values were generally slightly higher than those provided in the datasheet but agreed within measurement uncertainties. Deviations might be explained by variation between individual cameras, by the photodiode response having changed slightly in the two years since its calibration, or by the photodiode being positioned slightly farther from the light source than the sensor surface. When estimating proto-Lightspeed's throughput, we used the datasheet QE curve.

\section{Instrument Software and Timing}
\label{sec:software}
\subsection{Exposure time calculator}
\label{sec:etc}
To allow for performance estimation of proto-Lightspeed and support community use, we have developed an exposure time calculator (ETC) that predicts the signal-to-noise ratio (SNR) that will be obtained for a specified source and observation parameters. The ETC is available for public use at \url{https://lightspeed-astro.github.io/etc.html}. This ETC is written in python and uses the \textit{synphot} \citep{synphot:2018} and \textit{matplotlib} \citep{Hunter:2007} python libraries. The ETC can be run in a web browser at \url{https://lightspeed-astro.github.io/etc.html}, and its source code is publicly available\footnote{\url{https://github.com/ChrisLayden/Lightspeed-ETC}}. The ETC models atmospheric transmission, telescope and instrument transmission, scintillation noise, background light including moonlight, optimal aperture selection, and all known detector effects at a temperature of $-20^\circ$C (including read noise, dark current, and QE losses from nonlinearity).

Scintillation noise is traditionally calculated using Young's empirical model \citep{young:1967}. This model predicts
\begin{equation}
    \sigma_{scint}/F\approx 0.09 D^{-2/3}X^{7/4}e^{-h/H}(2t_{exp})^{-1/2}
    \label{eq:young}
\end{equation}
\noindent{}for primary mirror diameter $D$ in cm, airmass $X$, elevation $h$ in meters, scale height $H=8000$~m, and exposure time $t_{exp}$ in seconds. However, this model does not account for variations due to site or observing conditions and is known to underestimate the median scintillation noise by a factor of $\approx 1.5$ \citep{osborn:2018}. Therefore, the ETC estimates the scintillation noise by scaling the result of Eq.~\ref{eq:young} by this factor. The ETC does not yet account for noise from precipitable water vapor (PWV), which should be minimal for almost all of proto-Lightspeed's visible-wavelength observations.

The telescope transmission is calculated assuming $90\%$ reflectivity at each of the three mirror surfaces and a central obscuration diameter ratio of $29\%$, yielding a total throughput of $66.8\%$. The sky background in the $V$ band for given object and moon positions is calculated by summing Eqs. 2 and 15 in \cite{krisciunas:1991}. For observations at other wavelengths, the $V$ band brightness found in this manner is scaled using the background light spectrum modeled for Cerro Paranal, at similar conditions to LCO \citep{paranal:2013}. An optimal aperture is selected by iteratively adding one pixel to the aperture at a time until the SNR stops increasing. The ETC is easily configurable to add other image sensors or telescopes and study their performance. Section~\ref{sec:phot_prec} demonstrates the agreement achieved between this ETC and proto-Lightspeed observations.

\subsection{Instrument control software}
\label{sec:gui}
All of proto-Lightspeed's functionalities are controllable through its graphical user interface (GUI). This GUI, written in Python, runs on the proto-Lightspeed control computer, which operates using Ubuntu 22.04. The source code for this GUI is available upon request.
the GUI communicates with the Clay Telescope Control System (TCS) computer to display and save telescope state variables and allow for small coordinated offsets.

All images are saved as flexible image transport system (FITS) files. One header-data unit (HDU) of each file is an Image HDU, with a data cube containing multiple frames. If ``frame bundling" is selected in the GUI, multiple exposures will be concatenated into each frame in the data cube. This is useful for very fast windowed imaging, as it allows for fewer frames in each data cube, resulting in more efficient data writing. Each FITS file also contains a Table HDU containing time stamps corresponding to the end of each frame.

\subsection{Timing Calibration}
\label{sec:timing_cal}
Absolute timing is a critical capability of proto-Lightspeed, enabling several of its key science cases. Many ground-based optical instruments have historically not treated absolute timing robustly, often conflating absolute timing accuracy with relative timing precision, or relying on NTP-disciplined system clocks without accounting for latencies introduced by camera triggering and detector readout. In proto-Lightspeed, absolute timing is achieved using a dedicated Meinberg GPS180PEX PCI-Express card, which both triggers the camera via a pulse-per-second (PPS) signal and time-tags a TTL output from the camera corresponding to the end of detector readout.

Laboratory tests using a pulsed LED triggered on the PPS signal demonstrate that this configuration allows the mid-exposure time to be recovered with an accuracy better than 50~µs; this limit was set by the uncertainty in the LED rise time rather than by the timing system itself. While we also record a redundant timestamp corresponding to the start of each exposure, we find that the end-of-readout timestamp is less sensitive to exposure-start jitter, which we estimate to be of order 10~µs, and therefore provides a more robust reference for absolute timing. Section~\ref{sec:timing_results} demonstrates that proto-Lightspeed achieves timing accuracy better than 30~µs on sky.

\subsection{Image Processing Pipeline}
\label{sec:pipeline}
Before scientific analysis, proto-Lightspeed images undergo bias subtraction, nonlinearity calibration, flat fielding, and World Coordinate System (WCS) solution (when appropriate). In cases where dark current is non-negligible (only long exposures of low-flux sources, e.g., narrowband imaging), subtraction of a calibrated dark frame precedes flat fielding. As mentioned in Sec.~\ref{sec:read_noise}, we are developing a step following bias subtraction to take advantage of the sensor's DSERN to perform photon number resolving. proto-Lightspeed's field of view ($1'$) poses a challenge for astrometric plate solving. At present, we derive the WCS solution for proto-Lightspeed images using the python package \textit{twirl} \citep{2010AJ....139.1782L,2022MNRAS.509.4817G}, which is effective for fields of this size. We use a custom aperture photometry pipeline to conduct comparison photometry and generate light curves. Before proto-Lightspeed becomes available as a PI instrument in 2026b, its image processing pipeline will be made publicly available.

\section{Instrument Commissioning}
\label{sec:performance}
proto-Lightspeed was first commissioned on the Clay telescope on September 11--14, 2025. In this initial run, proto-Lightspeed lacked some components described in Secs.~\ref{sec:optical_design}~and~\ref{sec:mech_design} , including the Birger RF controller and the instrument cooling system, and the computer was mounted directly on the instrument. These components and fiber connectivity through the telescope were added prior to proto-Lightspeed's second run on the Clay Telescope from December 15--19, 2025. Using data from these runs, we measured the throughput of the instrument, demonstrated that its re-imaging optics did not degrade the telescope's image quality, confirmed the instrument's photometric performance and timing accuracy, and began performing high-impact science.

\subsection{RF lens complications}
During commissioning, we faced complications with the Canon RF lens used for re-imaging. While these complications did not noticeably affect proto-Lightspeed's image quality or capabilities, they made instrument operation and analysis of commissioning data slightly more cumbersome. We will resolve these complications before proto-Lightspeed becomes a PI instrument, but we mention them here to note their effects on the commissioning data and to inform others developing COTS imaging solutions.

First, we expected that the Birger RF controller, when powering the RF lens, would stabilize its internal optical components. When an RF lens is unpowered or when image stabilization is disabled, some such components are free to move radially as the lens rotates about its optical axis. This is not true of Canon EF lenses, for which stabilizing elements are held in place when the lens is unpowered. Unexpectedly, the image stabilization function of the Birger controller did not work during commissioning. This caused proto-Lightspeed's field to occasionally (at intervals of 5--60 minutes)  translate across the sensor in fast, discrete shifts, particularly during large changes in rotator angle.

\begin{figure}
    \centering
    \includegraphics[width=0.95\linewidth]{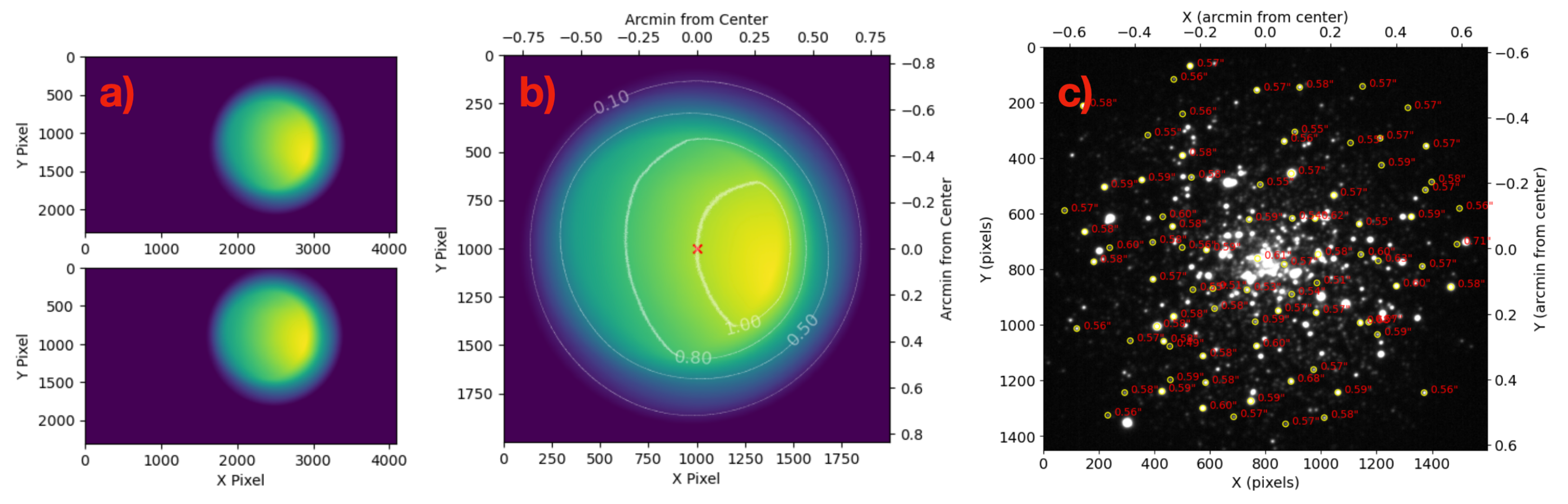}
    \caption{\textbf{a)}: Flat fields taken before and after a translation caused by the lack of image stabilization in the RF lens. \textbf{b)}: A flat field demonstrating vignetting caused by the COTS re-imaging optics. The asymmetry is likely caused by misalignment, again from the lack of image stabilization. Contours show the illumination relative to the geometric center of the field, with a maximum value of 1.15 reached slightly to the right of the center. \textbf{c)} Image quality measured across the field, using globular cluster M30. Select stars are labeled with their measured PSF FWHM, in arcseconds. Seeing-limited imaging is maintained across the field.}
    \label{fig:vignetting}
\end{figure}

Figure~\ref{fig:vignetting}a demonstrates the position of the field on the sensor before and after such a shift. In the maximally demagnified setting, the full $1'$ field remains on the sensor regardless of such translations, and the image quality is not degraded by these shifts. However, when reading out a small subarray of the sensor, the source may move outside of the subarray, requiring the observer to adjust the subarray. For long exposures, a mid-exposure translation will not be noticed until after the exposure is complete, resulting in a lost frame. In processing, frames may need to be shifted into alignment (the translations do preserve rotation) before co-addition or aperture photometry. The unstabilized (and, therefore, slightly misaligned) components in the RF lens also cause asymmetric vignetting of the field, as shown in Fig.~\ref{fig:vignetting}b. We will address this stabilization issue either with updated Birger controller firmware or by controlling the RF lens by wiring it to a separate camera body.

The second complication arises from an internal infrared LED within the RF lens. When the lens is powered on, this LED contributes a significant amount of stray light (much brighter than the sky background) at the focal plane. During commissioning, we opted not to power the Birger controller---and, thereby, the RF lens---to avoid this stray light and because image stabilization did not work. We will resolve the LED stray light issue either with lens disassembly and wire-snipping or an internal infrared blocking filter.

\subsection{Measured throughput}
\label{sec:throughput}
We probed the throughput of proto-Lightspeed by observing a field centered at $\alpha=$ 06:40:54.08, $\delta=-$19:35:15.8 in $g'$, $r'$, and $i'$ on Dec. 17, 2025. This field was chosen because it was at unit airmass and was observed by SDSS in these colors \citep{almeida:2023}, allowing for zero point and color transformation calculations. We recorded at least three exposures of this field with the $g'$, $r'$, and $i'$ filters using exposure times such that the brightest star yielded $\approx 50\%$ of saturation in its brightest pixel. We applied bias subtraction, nonlinearity calibration, flat fielding, and background subtraction to these exposures. We matched ten stars across the field to stars with known SDSS $g', r', i'$ magnitudes. We calculated the instrumental magnitudes $g'_{inst}, r'_{inst}, i'_{inst}$ for each star as $-2.5\log_{10}(F/G/t_{exp})$, where $F$ is the bias- and background-subtracted star brightness, in ADU, $G=8.9\,\mathrm{ADU/e}^-$ is the detector conversion gain, and $t_{exp}$ is the exposure time. Using these data, we performed least-squares fitting to the equations

\begin{align}
g'_{SDSS} &= g'_{inst} + ZP_{g'} + C_{g'}(g'_{SDSS}-r'_{SDSS}) \label{eq:gprime}\\
r'_{SDSS} &= r'_{inst} + ZP_{r'} + C_{r'}(g'_{SDSS}-r'_{SDSS}) \label{eq:rprime}\\
i'_{SDSS} &= i'_{inst} + ZP_{i'} + C_{i'}(r'_{SDSS}-i'_{SDSS}) \label{eq:iprime}
\end{align}

\noindent{}to determine the instrumental zero points $ZP_{g'},ZP_{r'},ZP_{i'}$ and color transformation terms $C_{g'},C_{r'},C_{i'}$ for converting to the SDSS color system. We report the median zero point for each filter, the overall system throughput inferred from these zero points, and the measured color terms in Table~\ref{tab:zps}. The relatively large color transformation terms for $g'$ and $i'$ are not unexpected. For example, because the throughput of the re-imaging optics outside of visible wavelengths is poor, objects will appear fainter in $i'$ than expected in the SDSS color system; thus, $C_{i'}$ is negative. $C_{g'}$ is likely measurably positive because the Baader $g'$ filter has better blue throughput than the SDSS $g'$ filter. For precise color photometry, more accurate color transformations should be derived using a significantly larger sample of stars with more diverse colors.

We stress that due to the detector nonlinearity and the field vignetting, the measured throughput values are accurate for the geometric center of the field for observations in the linear regime of the detector ($\gtrsim 100\,\mathrm{e}^-$/pix). When the image stabilization issue is resolved, we anticipate that the vignetting will be symmetric and improved. This effect largely explains the slightly lower than predicted throughput: the maximum of the flat field delivers $\approx 15\%$ higher throughput than the field center, to which the zero points are referenced.

\begin{table}[]
    \centering
    \begin{tabular}{c|c|c|c|c|c}
         Filter & Theoretical ZP & Measured ZP & Theoretical & Measured & Color Term \\
          & (AB Mag) & (AB Mag) & Throughput & Throughput & (to SDSS) \\
         \hline
         Baader g' & 27.8 & $27.6\pm0.1$ & 22\% & $19\pm2\%$ & $C_{g'}=0.14\pm0.04$ \\
         Baader r' & 27.3 & $27.2\pm0.1$ & 23\%& $21\pm2\%$ & $C_{r'}=0.00\pm0.03$ \\
         Baader i' & 26.0 & $26.0\pm0.1$ & 7\%& $7\pm1\%$ & $C_{i'}=-0.11\pm 0.03$
    \end{tabular}
    \caption{Zero points and throughput values predicted and measured for proto-Lightspeed. Theoretical values are calculated using the exposure time calculator described in Sec.~\ref{sec:etc}. Throughput values are calculated relative to a perfect 6.5~m-diameter reflector. Color transformation terms, defined in Eq.~\ref{eq:gprime}-\ref{eq:iprime}, are also given for conversion to the SDSS color system.}
    \label{tab:zps}
\end{table}

\subsection{Measured image quality}
\label{sec:image_quality}
To confirm that the COTS re-imaging components do not degrade the image quality delivered by the Clay telescope, we repeatedly compared the seeing measured by the telescope to the PSF size across proto-Lightspeed's field. Across both commissioning runs, proto-Lightspeed achieved seeing-limited observations across the field in the open, $g'$, $r'$, and $i'$ filters, although re-focusing was occasionally required at few-hour intervals. Figure~\ref{fig:vignetting}c shows globular cluster M30, as imaged by proto-Lightspeed in $r'$ with a co-added stack of 500 exposures, each of duration 30~ms, under relatively good seeing conditions in September 2025. After applying the processing pipeline to this image, we extracted the $\approx200$ brightest sources and measured their PSF FWHM via Gaussian fitting. We found a median PSF FWHM of $0.58''$, consistent with the seeing and with little to no variation across the field.

We found no measurable image distortion across proto-Lightspeed's field, as expected for its modest field of view. To probe field distortion, we matched the same processed and stacked image of M30 to the \textit{Gaia} star catalog \citep{gaiaDR3:2023}, using the WCS solution found using \textit{twirl} and matching stars within a search radius of $0.5''$. For matched stars, we calculated the separation vector between the WCS-determined and Gaia-measured sky positions. Across the field, the absolute separation was $\approx 110\pm70$~mas. Stars at larger radii from the center of the field did not exhibit larger separations, which would have indicated distortion: the $\approx 300$ matched stars $>0.5'$ from the center also had a mean absolute deviation of $\approx 100\pm70$~mas. Observing the separation vectors by eye across the field revealed no clear pattern in the directions of the separation vectors. For high-precision astrometry, further studies are warranted to probe the residual separations between measured and cataloged star positions, which may be caused by uncertainties in Gaussian centroid fitting.

\begin{figure}
    \centering
    \includegraphics[width=0.95\linewidth]{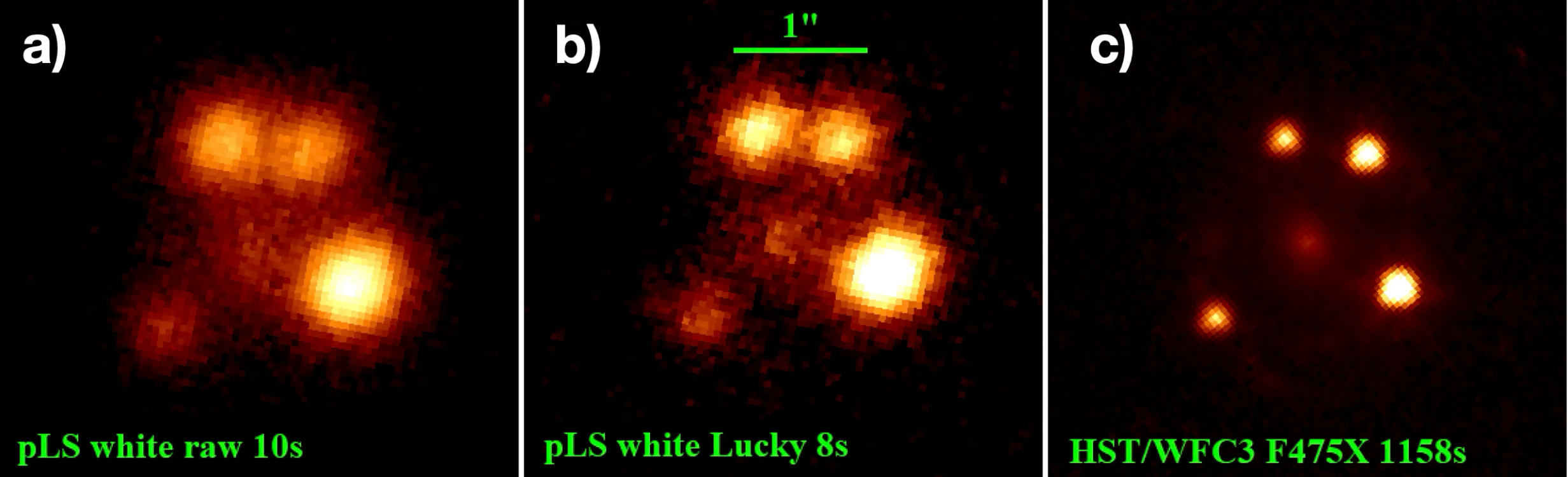}
    \caption{Quadruply lensed quasar DES J0420-4037 and its lensing galaxy. \textbf{a)} The system observed with a single 10~s proto-Lightspeed exposure in white light. \textbf{b)} A stack of the best $7\%$ of 600 frames with exposure time 200~ms improves the PSF FWHM from $0.51''$ to $0.37''$. \textbf{c)} For comparison, the system observed by the Hubble Space Telescope.}
    \label{fig:quad}
\end{figure}

Figure~\ref{fig:quad}a demonstrates how proto-Lightspeed's good image quality and LCO's excellent seeing conditions allow for the resolution of all four images in the gravitationally lensed quasar DES J0420-4037 \citep{shajib:2018}, as well as the lensing galaxy. This panel shows a single white-light 10~s exposure obtained in December 2025 with atmospheric seeing of $0.51''$, where even the closest images (separated by $\approx 0.6''$) are resolved. This verifies that proto-Lightspeed's image quality remained stable across the commissioning runs.

proto-Lightspeed can even deliver image quality slightly better than the atmospheric seeing by applying the technique of lucky imaging. In this technique, many short ($\lesssim 200$~ms) exposures are taken, and only the sharpest images are retained and stacked. Other instruments using smaller telescopes \citep{Law_2006} or operating at infrared wavelengths \citep{wong:2020} have used this technique to achieve diffraction-limited imaging from the ground. Due to the Clay telescope's large aperture relative to the size of the isoplanatic patch for visible light, proto-Lightspeed cannot reach this level of improvement \citep{Fried:78}. However, the image quality still generally improves by keeping and stacking short exposures that are least affected by atmospheric turbulence wavefront distortion. The central panel of Fig.~\ref{fig:quad} shows an improved image of DES J0420-4037 found by stacking the best $\approx 7\%$ of frames with exposure time 200~ms from a total of 120~s of observation. This stack provides a PSF FWHM of $0.37''$, allowing for clearer resolution of the lensing galaxy and better image localization. Thus, proto-Lightspeed can benefit from lucky imaging, and it easily delivers the frame rates necessary to do so. We have not yet evaluated proto-Lightspeed's speckle imaging capabilities.

\subsection{Measured timing accuracy}
\label{sec:timing_results}
\begin{figure}
    \centering
    \includegraphics[width=1.0\linewidth]{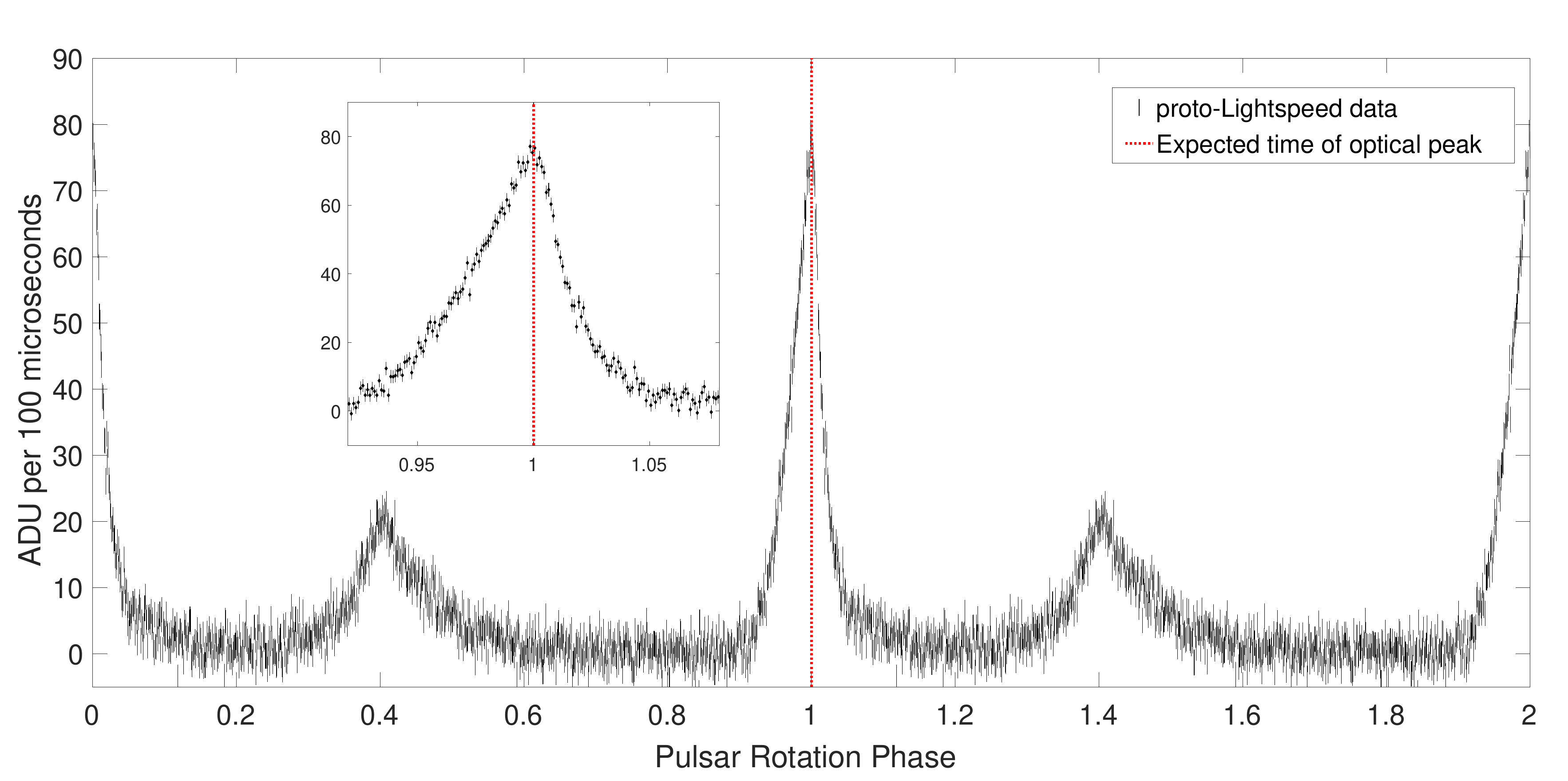}
    \caption{
    Optical light curve of the Crab pulsar obtained with proto-Lightspeed over a total integration time of $300\,\mathrm{s}$, sampled at $9259\,\mathrm{Hz}$ and phase-folded on the Crab’s rotational ephemeris.
    The red dotted line marks the expected phase of the optical main pulse, computed using the December 2025 radio ephemeris from the Jodrell Bank Observatory \citep{lyne:1993} and applying the known $\approx 255$~µs optical--radio pulse offset \citep{oosterbroek:2008}.
    Using proto-Lightspeed GPS-synchronized timestamps, we measure the maximum of the optical peak to be offset from the expected phase by $5\pm7$~µs in our December 17 data, well within the reported uncertainty of the optical--radio offset of $\approx 21$~µs \citep{oosterbroek:2008}.
    }
    \label{fig:crab}
\end{figure}

We verified the absolute timing performance on sky using observations of the Crab pulsar obtained during our December commissioning run (see Fig.~\ref{fig:crab}), when the GPS system was fully locked throughout the night. Figure~\ref{fig:crab} shows the optical light curve phase-folded on the Crab’s rotational period. We determine the time of pulse maximum by fitting a Gaussian to the phase-binned data in an interval of 0.01 of phase around phase=0 and derive uncertainties from the covariance matrix. Using a contemporaneous radio ephemeris from Jodrell Bank \citep{lyne:1993} and applying the known $\approx 255$~µs delay between the optical and radio pulses \citep{oosterbroek:2008}, we find that the peak of the optical main pulse occurs within $5\pm7$~µs of the expected phase. This offset is well within the reported uncertainty of the intrinsic optical--radio lag of $\approx21$~µs, indicating that the absolute timing accuracy of proto-Lightspeed is $\leq30$~µs, with an uncertainty intrinsic to the timing calibrators we have thus far used.

While such absolute timing precision may be unnecessary for long-period variables, it is essential for studies of pulsars—particularly millisecond pulsars, which are accessible with proto-Lightspeed—as well as for measuring wavelength-dependent lags in flaring X-ray binaries, searching for optical counterparts to FRBs, and studying other compact object systems. As an illustrative example, observations of the ultracompact binary ZTF~J1539+5027 with HiPERCAM achieved mid-eclipse timing precision of $8\,\mathrm{ms}$ despite a relatively long orbital period of $6.91$ minutes and exposure times of $3\,\mathrm{s}$. In this case, the timing precision is not limited by the exposure duration or orbital period, but instead by the extremely sharp ingress and egress of the eclipses, which allow the mid-eclipse time to be measured with very high precision. This demonstrates that even for longer-period systems observed with comparatively long exposures, robust absolute timing remains essential when the astrophysical signal itself permits precise temporal localization. Absolute mid-exposure timing at the sub-millisecond level is therefore a significant strength of proto-Lightspeed. We further plan to investigate row-level time-tagging of the detector readout, which could enable µs-level reconstruction of the effective mid-exposure time for individual pixels, even when operating at substantially slower frame rates.

\subsection{Measured photometric precision}
\label{sec:phot_prec}
\begin{figure}
    \centering
    \includegraphics[width=0.55\linewidth]{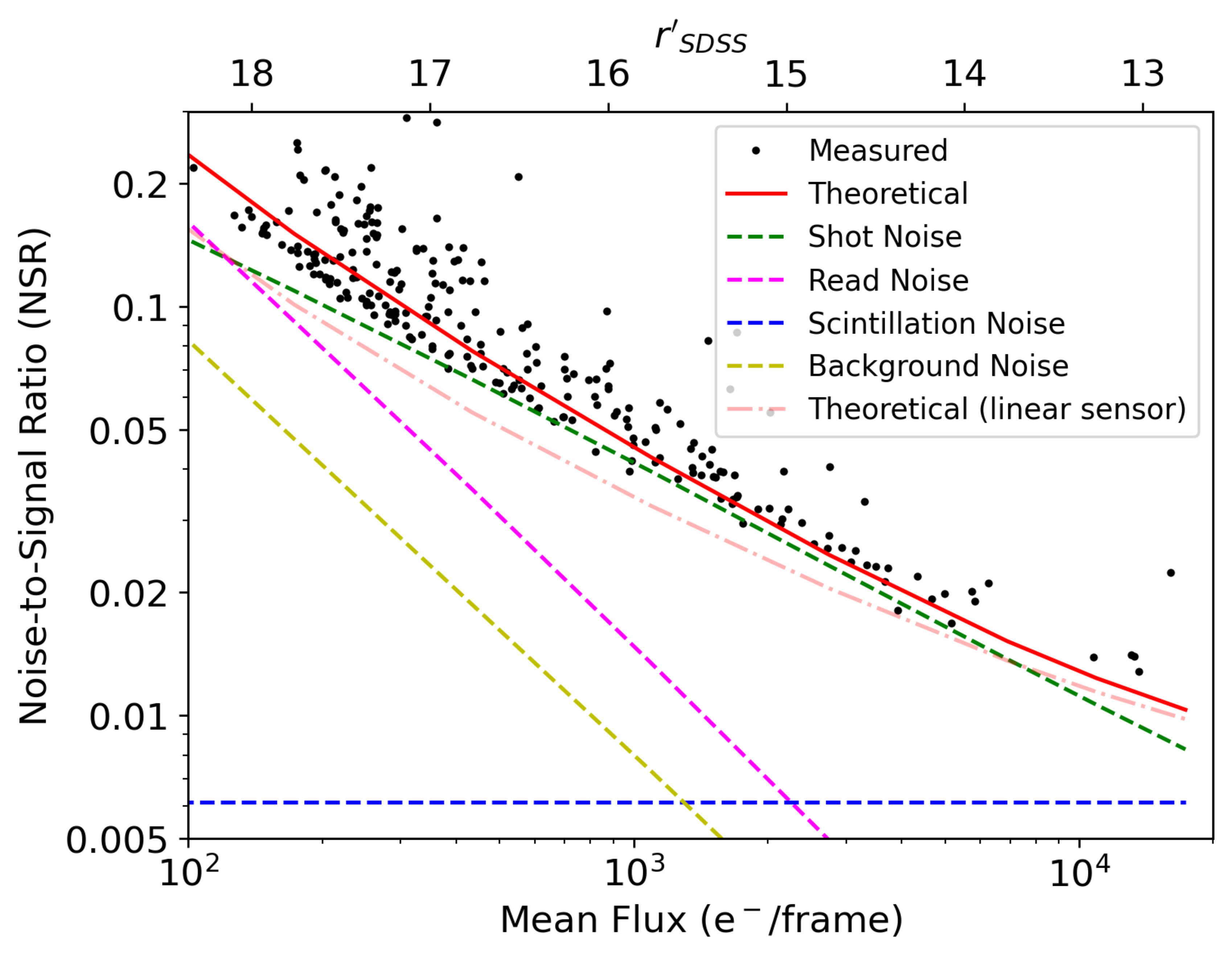}
    \caption{Noise-to-signal ratio (NSR) measured for sources identified in globular cluster M30 (black points), compared to theoretical predictions from the ETC (solid red curve). M30 was observed in $r'$ for 15~min at an exposure time of 30~ms. The total theoretical noise is the quadrature sum of source shot noise (green dashed), read noise (magenta dashed), scintillation noise (blue dashed), sky background noise (yellow dashed), and dark current noise (negligible). The theoretical noise that would be obtained if the sensor exhibited perfectly linear response is shown by the red dash-dotted curve. For small NSR values ($\lesssim 0.2$), $\sigma_{mag}\approx 1.086 \times$NSR. Instrumental magnitude is calculated as $-2.5\log_{10}{F/t_{exp}}$.}
    \label{fig:phot_prec}
\end{figure}

To confirm our understanding of proto-Lightspeed's noise properties, we analyzed the same exposures used to probe the image quality---500 images of globular cluster M30 with exposure time 30~ms in the $r'$ filter. As this yields a total duration of 15~s, faster than typical stellar variability timescales, any fluctuations in brightness across these frames are attributed to noise. After identifying $\approx 600$ stars in the field, we found 255 stars separated from any other stars by at least $1.2''$. We measured the average flux and standard deviation of the flux for all such stars across the 500 exposures, using a fixed aperture radius of $0.75''$. We used annuli with fixed inner and outer radii of $0.90''$ and $1.25''$ for median background estimation and subtraction. Figure~\ref{fig:phot_prec} shows the noise-to-signal ratio (NSR; the standard deviation divided by the mean) for each star as a function of stellar magnitude $r'_{SDSS}$ (found using the zero point from Sec.~\ref{sec:throughput}). We also found the total NSR expected as a function of stellar magnitude using the proto-Lightspeed exposure time calculator (red curve in Fig.~\ref{fig:phot_prec}), as well as the noise components contributing to this total (dashed curves). The measured average background level ($\sim 60\,\mathrm{e}^-$/star/frame) was well matched by the ETC prediction upon specification of the source airmass (1.01), moon phase (80\% illumination), and moon position (on the horizon). As discussed in Sec.~\ref{sec:ptc}, some noise components must be adjusted from the typical astronomical scaling due to the nonlinear sensor response that is calibrated away. The ETC automatically applies this rescaling, accounting for the signal level in each pixel in a simulated PSF. Figure~\ref{fig:phot_prec} also shows the NSR that would be attained if the sensor behaved as an ideal linear manner (red dash-dotted line); the difference with the measured NSR is more pronounced for faint sources. Again, this effect is best interpreted as a signal-dependent QE.

When shot noise and read noise dominate, the measured noise agrees well with the theoretical noise values, demonstrating that we have a good understanding of the sensor performance. For bright sources, the NSR starts to become independent of flux, as expected when scintillation noise dominates. The NSR for the brightest stars is slightly greater than the expected value, which may be attributable to site or temporal variations in scintillation noise. More detailed studies are needed to better parametrize the functional dependence of the scintillation noise at the Magellan Telescopes, with different exposure times and observing conditions. With its capability for high frame rates over a modest field, proto-Lightspeed could perform these studies, which could benefit all instruments at the observatory.

\subsection{Scientific Highlights}
\label{sec:science_highlights}
In proto-Lightspeed's two commissioning runs to date, we have observed a diverse set of astronomical phenomena, from low surface brightness pulsar wind nebulae to ultracompact binaries to optical pulsars. Here we present a snippet of the science already delivered by proto-Lightspeed.

\begin{figure}[t!]
    \centering
    \includegraphics[width=0.98\linewidth]{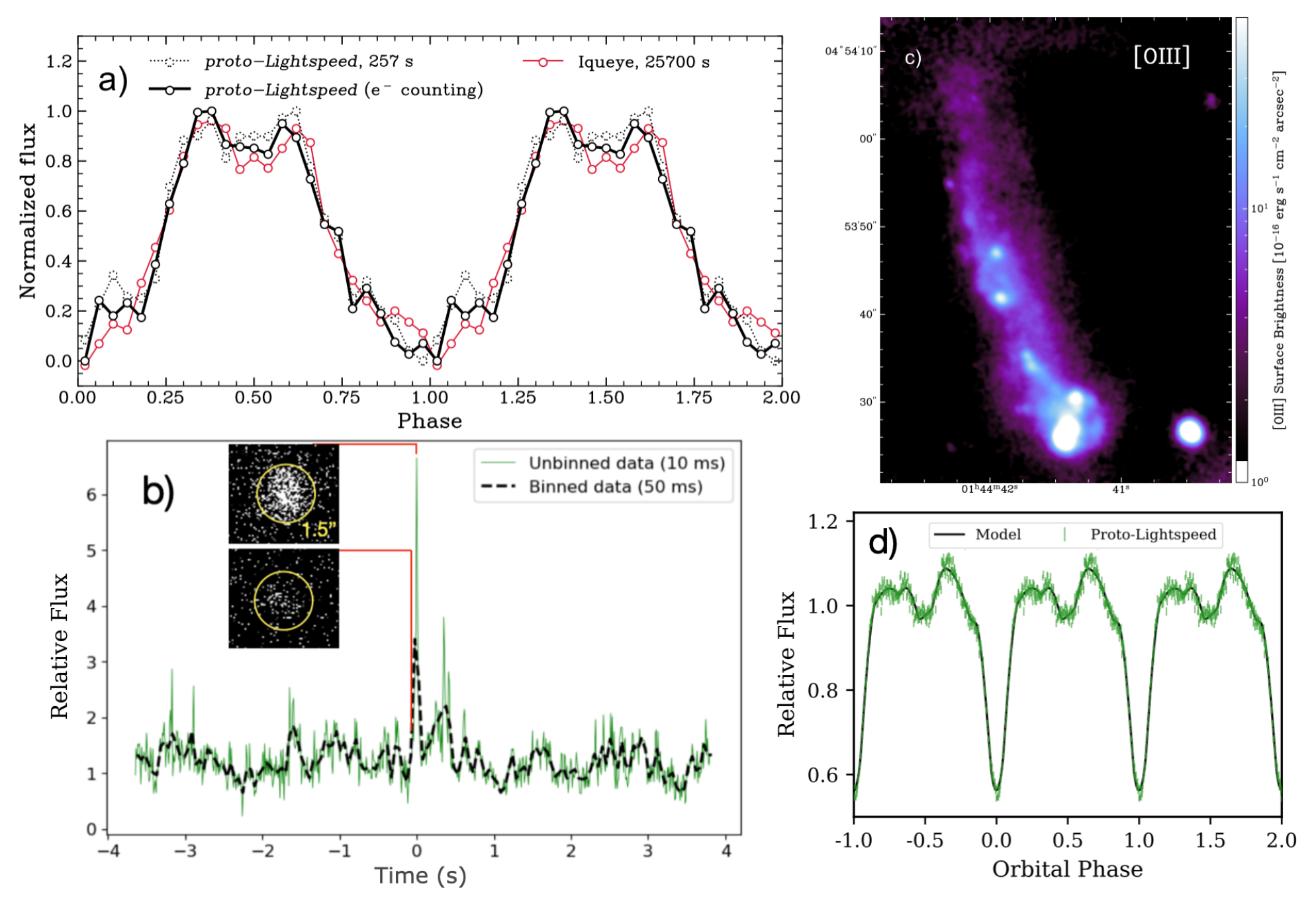}
    \caption{\textbf{a)} Phase-folded light curve of PSR B0540$-$69. Red represents the highest quality existing light curve, collected with Iqueye, while the black lines represent proto-Lightspeed's result with 1\% of the exposure time. The solid black line additionally employs an electron counting technique to reduce noise, which is enabled by proto-Lightspeed's deep sub-electron read noise. \textbf{b)} Light curve of BH XRB GX~339$-$4, sampled at 100 Hz (green curve). The light curve exhibits the highest amplitude short-timescale optical flares ever seen in an XRB \citep{berger:2026}. GX 339$-$4 has previously been observed optically at up to 20 Hz; the black dashed curve shows how flares are smoothed at this speed. Inset, back-to-back 10~ms exposures show the brightening is still not fully resolved. \textbf{c)} 15 minute [OIII] exposure of a nearby dwarf galaxy, demonstrating background-limited narrow-band imaging and high spatial resolution of the multi-phase ISM. \textbf{d)} Phase-folded light curve of ATLAS J013-4516, a mass-transferring white dwarf binary with orbital period 8.56~minutes, observed at cadence 1~s \citep{chickles:2026}. proto-Lightspeed provided insights into the geometry of this system and delivered its most precise eclipse timing to date.}
    \label{fig:science_highlights}
\end{figure}

PSR B0540$-$69, the ``Crab twin''  (hereafter B0540), is a 51~ms pulsar in the Large Magellanic Cloud (LMC) and the second brightest pulsar in the optical band after Crab itself. Its pulse profile is well studied at radio, X-ray, and $\gamma$-ray wavelengths. The highest quality V-band profile was constructed from 7.13 hours of Iqueye data collected in 2009 at the 3.58m NTT \citep{2011MNRAS.412.2689G}. In December 2025, proto-Lightspeed observed B0540 for 4.3 minutes in white light with 3~ms exposures. Figure~\ref{fig:science_highlights}a compares the proto-Lightspeed pulse profile (dotted black) to the NTT result (red, re-binned from 50 to 25 phase bins). Our data were phased using the ephemeris found by \cite{2024ApJ...962...92X}. We find a similar quality pulse in one one-hundredth of the observing time, although it is challenging to quantitatively compare the SNR of the proto-Lightspeed and Iqueye observations without knowledge of the noise estimation procedure employed for Iqueye.

The light curve can be further improved thanks to the deep sub-electron read noise of proto-Lightspeed's detector. As discussed in Sec.~\ref{sec:detector}, we are developing a photoelectron counting processing step for all ultra-quiet mode observations, taking advantage of the separation between peaks in the measured ADU distribution in each pixel, which represent individual electrons. 
The solid black line of Fig.~\ref{fig:science_highlights}a uses these peaks to reduce noise in the pulse profile, providing further SNR improvement, particularly at minimum. Bootstrapped errors were estimated by repeatedly re-sampling the data with replacement and measuring the per-bin standard deviation of the resulting light curves. These errors showed that this photon counting step reduced the noise by a factor of $\approx 1.8$ compared to the raw light curve. This result emphasizes the importance of deep sub-electron read noise, which will be critical for further characterizing faint and polarized optical pulsar light curves.

In September 2025, proto-Lightspeed observed outbursting black hole XRB GX~339 with a cadence of 10~ms. These observations showed that optical flaring in such systems can occur on time scales as short as 10~ms, faster and with significantly higher peaks than previously observed with longer exposures (see Figure~\ref{fig:science_highlights}b). Such fluctuations on timescales $t_{ms}$ (in milliseconds) probe a radius $r$ close to the innermost stable circular orbit (ISCO) scale, with $r/R_{\rm ISCO} \approx 2 (t_{\rm ms}/M_{10})^{2/3}$ for a black hole of mass $M_{10}\times 10M_\odot$. The implications of the fast flaring in GX~339 will be presented in more detail in an upcoming paper \citep{berger:2026}. Future simultaneous observations of outbursting XRBs with X-ray telescopes would uncover a wealth of information about the geometries of such systems, as proto-Lightspeed could deliver the most precise flare time delay measurements to date.

proto-Lightspeed's low read noise also provides an excellent opportunity for high-SNR narrow-band imaging, which is useful in many applications. One example is mapping the [OIII] and H$\alpha$ emission of nearby dwarf galaxies, taking advantage of Magellan's excellent angular resolution to study the multi-phase ISM. In December 2025, we collected a 15 minute [OIII] exposure of a nearby dwarf galaxy (DESI ID 39627905885538609, distance 19.7 Mpc) with proto-Lightspeed (Fig.~\ref{fig:science_highlights}c). The noise in this image is dominated by the sky background. Even for narrow-band H$\alpha$ imaging in the darkest conditions, the sky background contributes more to the noise budget than read noise as long as the exposure time is longer than 20~s.

Finally, proto-Lightspeed's observations of the ultracompact, disk-accreting AM Canum Venaticorum (AM~CVn) system ATLAS~J1013$-$4516 demonstrate the instrument's sensitivity to faint, rapidly varying sources. ATLAS~J1013$-$4516 is a mass-transferring white dwarf binary with a mean \textit{Gaia} magnitude of $G=19.51$ and an orbital period of only 8.56 minutes, placing it among the shortest-period disk-accreting binaries known \citep{chickles:2026}. In December 2025, proto-Lightspeed observed this system with 1 second integrations over several consecutive orbital cycles, sharply resolving the system's highly structured eclipse morphology in a regime inaccessible to synoptic surveys (see Fig.~\ref{fig:science_highlights}d). The high-cadence proto-Lightspeed light curve delivered the most precise individual eclipse timings for this system to date, illustrating the instrument's strength for studying orbital evolution. In addition, the detailed eclipse morphology resolved by proto-Lightspeed---including kinked ingress and egress features associated with the accretion stream and disk geometry---provided valuable input for light curve modeling \citep{chickles:2026}.

\section{Future plans}
\label{sec:future}
\begin{figure}[t!]
    \centering
    \includegraphics[width=0.9\linewidth]{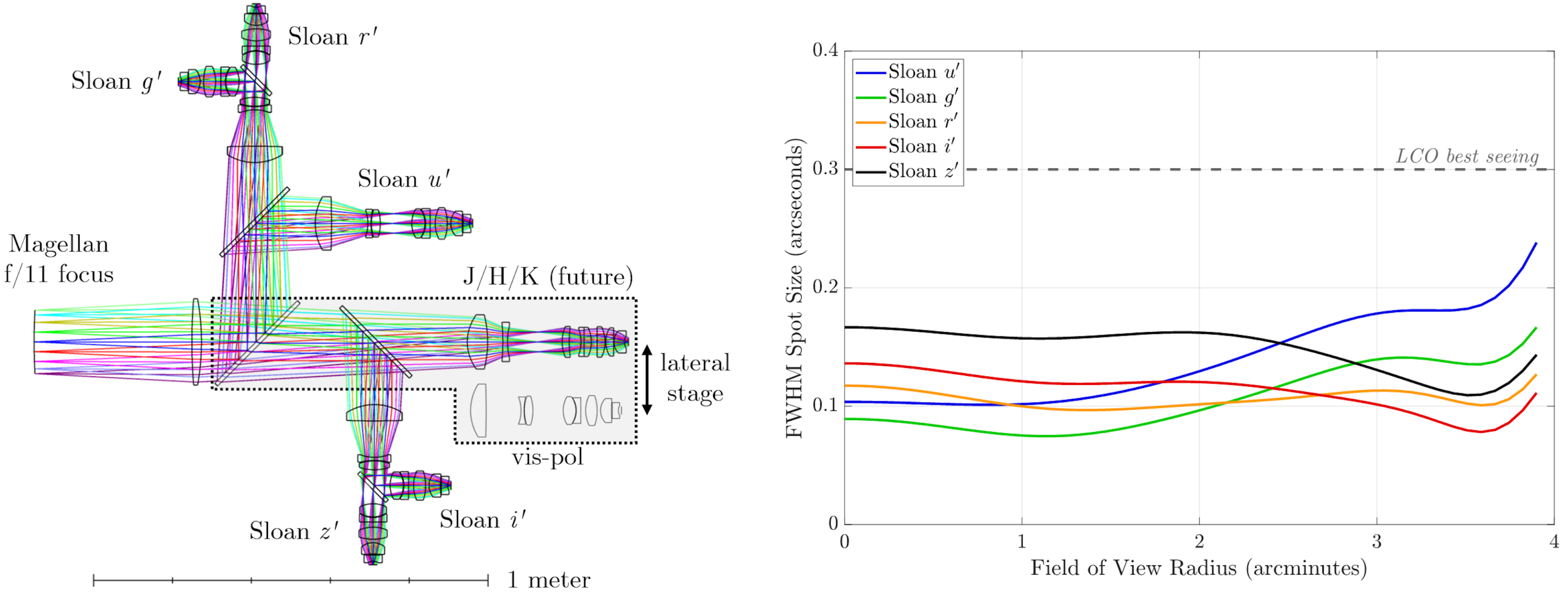}
    \caption{Left: Optical design of the full Lightspeed instrument, allowing for either five-channel $u',g',r',i',z'$ simultaneous imaging (with the potential for an additional infrared arm) or a single white light channel with polarimetric capabilities. Right: Designed image quality of Lightspeed's five optical channels.}
    \label{fig:full_lightspeed}
\end{figure}

We will conduct two final commissioning runs of proto-Lightspeed in March and May 2026, in which we will resolve lingering issues (including image stabilization) and possibly add new capabilities, including WeDoWo polarimetry and a motorized tip mount for narrow-band filters to study line emission from low-redshift galaxies. We may also evaluate proto-Lightspeed's speckle interferometry capabilities and develop a data reduction pipeline for such observations. proto-Lightspeed will be available as a PI instrument beginning in the 2026B observing cycle.

proto-Lightspeed serves as a prototype for a simultaneous multicolor imager, Lightspeed, intended to be a facility instrument on the Clay telescope. Figure~\ref{fig:full_lightspeed} shows the optical design of Lightspeed. Lightspeed will have five channels ($u',g',r',i',z'$), each hosting an ORCA-Quest 2 camera. Infrared light will pass straight through the dichroic filters, leaving the possibility for an additional photon-counting infrared arm, potentially employing an HgCdTe linear mode APD \citep{baker:2024}. An alternate ``VIS-POL" channel will allow for polarimetry and white-light imaging. Lightspeed will use custom re-imaging optics to deliver a plate scale of $0.1''$/pix and field of view of $7'\times4'$, without degrading the image quality provided by the telescope (see right panel of Fig.~\ref{fig:full_lightspeed}). These custom re-imaging optics will yield a significant increase in throughput compared to proto-Lightspeed, and the significantly larger plate scale will critically sample the PSF in good conditions and thereby decrease read noise even further for a given source. The 16-times larger field will allow for studies of large extended sources and would yield an order-of-magnitude increase in sensitivity to serendipitous occultation by small TNOs. The design specifications of Lightspeed are included as an option in the ETC (see Sec.~\ref{sec:etc}). Using these specifications, we predict that Lightspeed will achieve zero points of $u'=25.9$, $g'=28.3$, $r'=27.7$, $i'=27.3$, and $z'=26.4$.

Beyond Magellan, proto-Lightspeed also benefits the development of \textit{Cerberus}, a three-channel imager ($u'$, $g'$, $r'/i'/z'$) for the 200-inch Hale telescope at Palomar Observatory, slated for first light in late 2026 and intended to replace CHIMERA \citep{cerberus_mo_inprep}. \textit{Cerberus} will also use the ORCA-Quest 2 cameras with a similar planned plate scale of 0.13$^{\prime\prime}$/pix and a $9'\times4.5'$ field of view, allowing for significant overlap in control software and data reduction pipelines. Together, Lightspeed and \textit{Cerberus} will deliver high-speed multicolor imaging capabilities with DSERN CMOS detectors on large telescopes in both the Northern and Southern hemispheres.

\section{Conclusion}
\label{sec:conclusion}
proto-Lightspeed for the first time brings ultra-fast imaging and ultra-low read noise to the Magellan telescopes. It is therefore ideally suited to studies of rapidly varying phenomena or low surface brightness sources. proto-Lightspeed demonstrates the unique capabilities of DSERN CMOS image sensors but also highlights some of their present limitations---namely, small pixels and nonlinear response at low light levels. We address these limitations via re-imaging optics and calibration. Despite using COTS re-imaging components for rapid deployment, proto-Lightspeed delivers seeing-limiting image quality and deep imaging in the $g'$, $r'$, and $i'$ bands over an $\approx 1'$ field. proto-Lightspeed also provides absolute timing precision better than 50~µs, narrow-band imaging, and the potential for other capabilities including speckle interferometry and full-Stokes polarimetry.

\section*{Disclosures}
We declare no financial, commercial, or other conflicts of interest related to the research presented in this paper.

\section*{Code, Data, and Materials Availability}
The data generated and analyzed in this study are not publicly available due to their large volume. However, the code and a representative subset of the data used for this work are available from the corresponding author upon reasonable request.

\section*{Acknowledgments}
This work was funded in part by the MIT Kavli Research Investment Fund (grant MKI KRIF 2654352 and 2654373) and in part by Kavli Institute Collaboration Kickstarter (KICK) Grant PS-2025-GR-0251-3110. JGM gratefully acknowledges support from the Heising-Simons Foundation and the Pappalardo family through the MIT Pappalardo Fellowship in Physics.

We warmly thank the technical support staff at Las Campanas Observatory for their assistance in machining parts, interfacing with the telescope, troubleshooting issues, and nightly operations. MIT undergraduate Jonah Klein assisted with detector characterization. Post-baccalaureate Stephen Wang assisted with improving the speed of the application of the detector nonlinearity calibration.

Some software was developed with the assistance of AI-based tools, including Anthropic's Claude AI, OpenAI's ChatGPT, and GitHub Copilot. All software was reviewed and tested by team members prior to use. This research made use of Photutils, an Astropy package for detection and photometry of astronomical sources \cite{bradley:2024}.

Funding for the SDSS and SDSS-II has been provided by the Alfred P. Sloan Foundation, the Participating Institutions, the National Science Foundation, the U.S. Department of Energy, the National Aeronautics and Space Administration, the Japanese Monbukagakusho, the Max Planck Society, and the Higher Education Funding Council for England. The SDSS Web Site is \url{http://www.sdss.org/}.

The SDSS is managed by the Astrophysical Research Consortium for the Participating Institutions. The Participating Institutions are the American Museum of Natural History, Astrophysical Institute Potsdam, University of Basel, University of Cambridge, Case Western Reserve University, University of Chicago, Drexel University, Fermilab, the Institute for Advanced Study, the Japan Participation Group, Johns Hopkins University, the Joint Institute for Nuclear Astrophysics, the Kavli Institute for Particle Astrophysics and Cosmology, the Korean Scientist Group, the Chinese Academy of Sciences (LAMOST), Los Alamos National Laboratory, the Max-Planck-Institute for Astronomy (MPIA), the Max-Planck-Institute for Astrophysics (MPA), New Mexico State University, Ohio State University, University of Pittsburgh, University of Portsmouth, Princeton University, the United States Naval Observatory, and the University of Washington.

This work has made use of data from the European Space Agency (ESA) mission \textit{Gaia} (\url{https://www.cosmos.esa.int/gaia}), processed by the \textit{Gaia} Data Processing and Analysis Consortium (DPAC, \url{https://www.cosmos.esa.int/web/gaia/dpac/consortium}). Funding for the DPAC has been provided by national institutions, in particular the institutions participating in the \textit{Gaia} Multilateral Agreement. 

This research made use of the monthly radio ephemerides provided by the Jodrell Bank Centre for Astrophysics
at \url{http://www.jb.man.ac.uk/~pulsar/crab.html}.

\bibliography{references}{}
\bibliographystyle{aasjournal}

\end{document}